\documentclass[9pt,twocolumn]{article}
\usepackage[margin=0.75in]{geometry}
\usepackage{authblk}
\usepackage{lipsum} % for dummy text
\usepackage{abstract}
\usepackage{titlesec}
\usepackage{multicol}
\usepackage{fancyhdr}
\usepackage[sorting=none, style=numeric-comp]{biblatex}
\usepackage{textgreek}
\usepackage{wrapfig}
\usepackage{algorithm}
\usepackage{algpseudocode}
\usepackage[newcommands]{ragged2e}
\usepackage{graphicx}
\usepackage{amsmath}
\usepackage{amssymb}
\usepackage{caption}
\usepackage{subcaption}

\usepackage{tabularx}

\DeclareMathOperator*{\argmin}{arg\,min}

\title{Development of a Photon-Counting Deadtime Noise Model that Extends Dynamic Range and Resolution in Atmospheric Lidar}
\author[1*]{Grant J. Kirchhoff}
\author[2]{Matthew Hayman}
\author[3]{Willem J. Marais}
\author[1]{Jeffrey P. Thayer}
\author[4] {Rory A. Barton-Grimley}
\affil[1]{University of Colorado Boulder, Ann and H.J. Smead Aerospace Engineering Sciences Department, Boulder, CO, USA 80309}
\affil[2]{NSF National Center for Atmospheric Research, Earth Observing Lab, Boulder, CO, USA, 80303}
\affil[3]{University of Wisconsin Madison, Space Science \& Engineering Center, Madison, WI, USA, 53706}
\affil[4] {NASA Langley Research Center, Hampton, VA, USA, 23681}
\affil[*]{Corresponding author: grant.kirchhoff@colorado.edu}

\date{}

\begin{document}

% \date{}

\twocolumn[
\begin{@twocolumnfalse}
\maketitle

\begin{abstract}
\noindent This work derives and validates a noise model that encapsulates deadtime of non-paralyzable detectors with random photon arrivals to enable advanced processing, like maximum-likelihood estimation, of high-resolution atmospheric lidar profiles while accounting for deadtime bias. This estimator was validated across a wide dynamic range at high resolution (4 mm in range, 17 ms in time).  Experiments demonstrate that the noise model outperforms the current state-of-the-art for very short time-of-flight (2 ns) and extended targets (1 $\mu s$).  The proposed noise model also produces accurate deadtime correction for very short integration times.  This work sets the foundation for further study into accurate retrievals of high flux and dynamic atmospheric features, e.g., clouds and aerosol layers. 
\end{abstract}
\vspace{1em}
\end{@twocolumnfalse}
]

\section{Introduction} \label{sec: intro}

Atmospheric lidars capture high-resolution intensity profiles in the atmosphere that are used to derive vertically resolved estimates of geophysical variables, such as water vapor, cloud and aerosol properties \cite{TOPAZ-ground, HSRL-2, MPD, HALO}. These observations advance the understanding of atmospheric processes and highlight the role of atmospheric lidar measurements in areas like numerical weather prediction, air quality forecasts, and climate model projections \cite{ESDS, NASA-PBL}. Accurate estimates of photon fluxes are essential to derive quantitative variables from lidar backscatter signals. However, errors caused by detector limitations can undercut their accuracy. For systems that employ photon-counting detectors (highly sensitive to low light levels), high photon flux levels on the detector result in nonlinear biases that propagate as errors in the retrieved data product \cite{Muller}. 

Excess photon flux poses an issue to photon counting because of a detector characteristic known as deadtime, where the detector undergoes a period of inactivity (or ``dead'' time) following a detection event, resulting in subsequent photons failing to register as detections. Deadtime behavior can be described in two ways: (1) extended deadtime - where incident photons during the deadtime period can extend the deadtime, and (2) non-extended deadtime - where the deadtime is fixed after it has been triggered \cite{Muller}. This work will focus on non-extended deadtime because non-paralyzable detectors (such as actively quenched single-photon avalanche diodes, or SPADs) are employed across many atmospheric lidar systems and exhibit this deadtime type \cite{SPAD}. This work does not apply to paralyzable detectors that exhibit extended deadtime, such as photomultiplier tubes. Hereafter, non-extended deadtime will be called ``deadtime'' for simplicity. 

Overcoming the limitation of deadtime bias is vital because deadtime imposes an upper limit on backscatter flux for photon counting, limiting the observability of atmospheric targets that exhibit high backscatter intensity (e.g., clouds, dense aerosol layers, smoke). Although analog detection systems work well in these flux regimes, the large variability and heterogeneity in backscatter flux in atmospheric targets often span a range of fluxes beyond the dynamic range of analog detectors. Hybrid-detection systems (that employ a combination of photon-counting and analog channels) \cite{Hybrid} or multi-gain systems (that employ multiple detectors with different gain or optical attenuation), e.g., \cite{Goldsmith, RAMSES, ALOMAR}, demonstrate some success in overcoming this limitation but stitching data from the different channels together without distortion is susceptible to error. For photon-counting systems, the ability to operate the detectors closer to their saturation limit would enhance their dynamic range and enable observations that extend from clear air to high backscatter targets, such as clouds. In hard-target lidar applications, techniques for deadtime-bias correction have been demonstrated for non-gated, hard-target ranging \cite{Rapp, Range-walk-error, Ye-range-walk-error, Hernandez-3D-lidar-image-construction}, but not yet for volumetric targets in atmospheric lidar, which is the subject of this paper. Some pseudo-volumetric applications avoid deadtime effects by utilizing time-gating \cite{Time-gating, Gated-viewing, Yang-Cloud}, in which the detector is only active for a short window of time per laser shot to block out saturation effects. However, extending this technique to volumetric targets in the atmosphere presents challenges because the location of the window in range needs to be swept across the body of the target, resulting in a scanning process slower than most meteorologically relevant atmospheric targets, e.g., cloud formation, based on observations reported in Ref. \cite{Hayman-AMS-2023}.  

Direct corrections for deadtime bias exist but are limited in their application to atmospheric lidar signals, necessitating a more advanced approach for processing signals impacted by deadtime bias. In Yang et al. \cite{Yang-deadtime}, conditional probability was leveraged to develop a direct correction for photon counts that accounts for correlated deadtime bias from preceding histogram bins. While an improvement over the traditional deadtime correction technique (see Sec. \ref{subsec: Muller approx}), this approach can only be applied to direct retrievals, limiting its application to advanced retrievals. Processing direct retrievals using the standard estimation technique (the flux estimate $\tilde{\lambda}$ is equal to the scaled counts in a histogram bin $\tilde{\lambda}=y/\mathcal{N}\Delta t$, where $y$ is the raw counts, $\mathcal{N}$ is the number of laser shots, and $\Delta t$ is the bin size) presents issues at high resolutions, where data in each bin becomes sparse and the estimate becomes very sensitive to shot noise. In Hayman et al. \cite{Hayman-2D-estimation}, the authors highlight how processing data at high resolutions can be enabled by maximum-likelihood estimation (MLE), where sparse regions are de-emphasized during the fitting routine by using the appropriate Poisson point-process model. The Poisson noise model assumes an ideal photon counter, excluding situations where the detector is pushed into nonlinear regimes, e.g., deadtime. In addition, they highlighted that atmospheric variability can be statistically significant below fractions of a second, suggesting that many atmospheric scenes may be undersampled. In Hayman et al. \cite{Hayman-AMS-2023}, it was suggested that averaging over temporally dynamic high flux regimes imparts previously unquantified errors into captured lidar signals. To understand these effects and how to mitigate them, it becomes necessary to process scenes at high resolutions similar to Ref. \cite{Hayman-2D-estimation} while accounting for detector deadtime. This study establishes a key step in this approach by developing a more accurate noise model that includes the effects of deadtime (which will be referred to as the "deadtime noise model"), thus enabling MLE to process high-resolution retrievals that are not possible through conventional direct estimation techniques. This new approach can facilitate the capture of fine features in atmospheric targets that exhibit high backscatter, such as clouds and aerosol layers.

In this paper, the deadtime noise model will be derived from first principles and validated on experimental data. The model will be evaluated at high range and temporal resolution across multiple orders of magnitude of photon fluxes that span the linear and nonlinear regimes of the detector. This will demonstrate its performance by comparing the experiment results against contemporary deadtime correction methods, while simultaneously identifying the model's limitations. To accomplish this objective, these tests will be performed by oversampling static targets at high resolution, a requirement for evaluating the noise model. By deriving and robustly validating this deadtime noise model through experimentation, this work establishes a critical foundation for future research efforts to understand and address errors in atmospheric lidar data imposed by high dynamic range and spatially heterogeneous targets. 

\section{Theory} \label{sec: theory}

\subsection{Limitations to the M\"uller Correction} \label{subsec: Muller approx}

Traditionally, in atmospheric lidar, deadtime bias has been compensated by applying the ``M\"{u}ller Correction'' (popularized by Donovan et al.) \cite{Donovan, Muller} as a direct correction to photon counts. This approach has limitations in that the M\"{u}ller Correction can only be applied to scenes that are approximately uniform (in range and time) over a wide sampling bin width. Although slowly varying scenes like this are encountered in atmospheric lidar (e.g., clear air), this constraint disqualifies targets that exhibit gradients in backscatter intensity, such as mixed aerosol layers and cloud edges, which are desired observables in the modeling community \cite{NASA-PBL}. 

This section introduces the approximations that are implicitly made when applying the M\"{u}ller Correction to deadtime-affected measurements and discusses the issues that arise when the correction is applied to improperly conditioned data. The M\"{u}ller Correction in its common form is written as
\begin{equation} \label{eq: Muller correction}
    \lambda=\frac{R}{1-R\tau},
\end{equation}
where $R$ is the measured photon-count rate, $\tau$ is the deadtime interval, and $\lambda$ is the estimate of the actual mean count rate \cite{Muller}. The (often overlooked) approximation that is made when applying this correction is that the measured count rate (calculated as $R=\langle N\rangle/\Delta t$, where $\langle N\rangle$ is the average number of photon counts in a bin and $\Delta t$ is the bin accumulation interval) represents an independent measurement of a uniform count rate within the bin. This concept can be organized into two related approximations that will be called the ``M\"{u}ller Approximations":

\begin{itemize}
    \item \textbf{Approximation \#1}: The photon flux during a bin interval $\Delta t$ is approximately constant.
    \item \textbf{Approximation \#2}: Each observation bin is unaffected by observations in other bins. This is generally valid when Approximation \#1 holds \textit{and} the bin time $\Delta t$ is much larger than the dead time $\tau$, i.e., $\Delta t\gg\tau$.
\end{itemize}

\begin{figure}[H]
\centering
\begin{subfigure}[t]{0.42\textwidth}
\centering
\includegraphics[width=\textwidth]{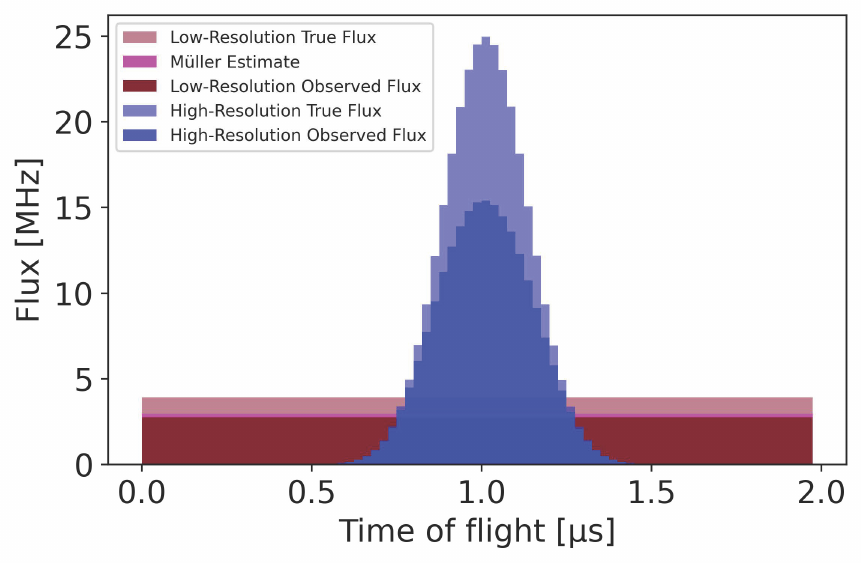}
         \caption{}\label{subfig: Muller histograms}
\end{subfigure}

\begin{subfigure}[t]{0.42\textwidth}
\centering
\includegraphics[width=\textwidth]{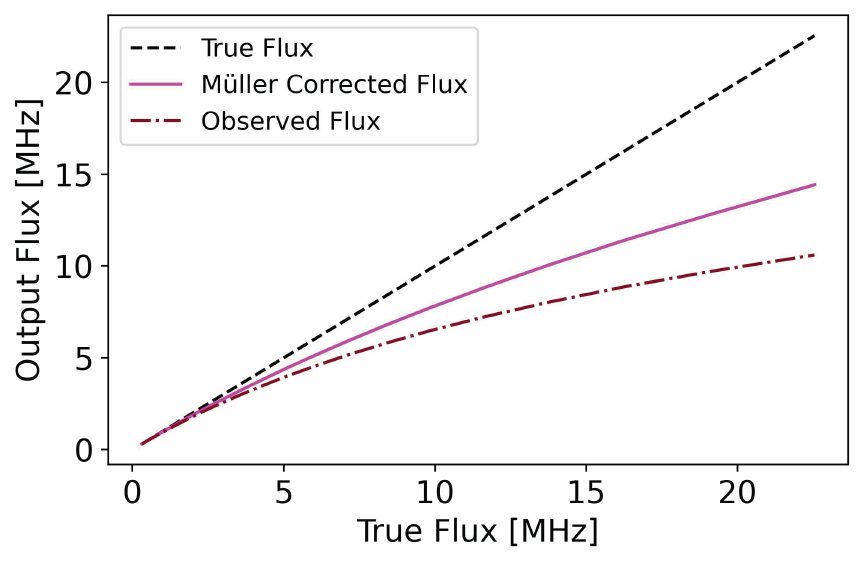}
         \caption{}\label{subfig: Muller curves}
\end{subfigure}
\caption{\textbf{Approximation \#1 violation}: (Top) Example of observed and true photon-count histograms of a measured narrow Gaussian pulse. The histograms were generated at fine bin widths $\Delta t_f$ (blue) and coarse bin widths $\Delta t_c$ (red). (Bottom) Interrogating the Gaussian pulse (with the same standard deviation as the left subfigure) and increasing the peak flux while integrating at the coarse bin width $\Delta t_c$. Approximation \#2 is satisfied only for sampling at coarse bin widths in this configuration.}
\label{fig: Muller correction test}
\end{figure}

The validity of Approximation \#1 is not determined solely by the histogram bin width of the data but also by the structure of the volumetric target under investigation. In Ref. \cite{Donovan}, Donovan et al. explicitly make Approximation \#1 (homogeneous flux) when deriving the M\"{u}ller Correction. Fig. \ref{fig: Muller correction test} demonstrates how applying the correction to scenes incompatible with Approximation \#1 (in this case, a nonhomogeneous flux) results in an erroneous correction. In Fig. \ref{subfig: Muller histograms}, two distinct simulated acquisitions of the same narrow target with a Gaussian backscatter profile of standard deviation $\sigma=10\tau$, where $\tau=25$ ns is the deadtime interval. The blue-hue histograms represent the profile acquired at high resolution (fine bin width $\Delta t_f=25$ ns), while the red-hue histograms represent the profile acquired at low resolution (coarse bin width $\Delta t_c=2$ $\mu$s). At both resolutions, the measured flux is biased low by deadtime. Note that the M\"{u}ller estimate for the coarse measurement does not recover the true flux accurately. In this figure, the M\"{u}ller estimate was not applied to the high-resolution measurement because that would violate Approximation \#2 (since $\Delta t_f=\tau$), which will be discussed separately. Fig. \ref{subfig: Muller curves} displays the results when repeated coarse measurements are made while gradually increasing the amplitude of the Gaussian target. At the higher fluxes, the M\"{u}ller Correction underestimates the actual flux, an error caused by violating Approximation \#1. Although clear air and aerosol structures with low variability typically satisfy Approximation \#1, other scenes that exhibit heterogeneity (particularly near clouds) rarely conform to this requirement. Moreover, there is increasing evidence that using coarse resolutions that oversmooth the scene's heterogeneity can degrade the accuracy of derived products from remote sensing data \cite{Alkasem, Arola}. 

An intuitive solution to satisfy Approximation \#1 while observing a nonhomogeneous flux is to sample with narrow bin widths that match the spatial frequency in the scene (to approximate uniform fluxes within each bin), like the high-resolution measurement in Fig. \ref{subfig: Muller histograms}. Indeed, ultra-high-resolution measurements, on the scale of picoseconds, are now possible with the adoption of time-correlated single photon counting (TCSPC) \cite{RBG-TCSPC, T2, Hayman-2D-estimation} in combination with high-repetition-rate, short pulse-width lasers in atmospheric lidar. However, the narrow bin widths risk incompatibility with Approximation \#2, in which the deadtime length restricts the minimum allowable bin width (to reduce correlational effects between adjacent bins), an explicit requirement in Donovan et al. \cite{Donovan}. For example, a typical SPAD detector's deadtime interval can be 30 ns (or 4.5 m in range); thus, to satisfy Approximation \#2, the bin width $\Delta t$ would need to be greater than 300 ns (or 45 m). At the same time, the variability of atmospheric scatterers can occur on range scales finer than this, thus creating contradicting requirements between Approximation \#1 and \#2. Examples where the M\"{u}ller Correction is applied at high resolutions (thus violating Approximation \#2) are shown and discussed in Appendix \ref{subsec: Muller appendix}.

The paradox imposed by observing deadtime-affected, spatially varying scenes while simultaneously satisfying the M\"{u}ller Approximations necessitates a more robust solution to process high-flux scenes that exhibit uniformity and heterogeneity. This work introduces a novel deadtime noise model that enables MLE to recover accurate photon count rates without restrictions on resolution, such as being constrained by the M\"{u}ller Approximations or suffering from shot noise sensitivity when estimating flux from histograms at high resolution. This approach enables accurate and precise, high-resolution measurements in the presence of deadtime, enabling the capture of scenes that exhibit high backscatter variability. This solution can improve lidar data accuracy, increase data availability, and extend the dynamic range of photon counting for many atmospheric applications. The following section will provide details of this solution. 

\subsection{Photon Counting Model with Deadtime} \label{subsec: deadtime model}

\subsubsection{Deadtime Noise Model}

The ideal model for photon counting is a Poisson point process that is parameterized by the photon arrival rate $\lambda(t)$ using the continuous-time dimension $t$. Specifically, this process is defined as a nonhomogeneous Poisson point process, in which $\lambda(t)$ is not necessarily constant in time (or equivalently range), e.g., heterogeneous volumetric target. From Ref. \cite{Snyder}, this process is defined by the following probability distribution:
\begin{equation} \label{eq: Poisson model}
    P\left(\{T_n=t_n\}_{n=1}^{N}\right)=\exp\left(-\int_{0}^{t_N}\lambda(t')dt'\right)\prod_{n=1}^{N}\lambda(t_n),
\end{equation}
where $\lbrace T_n\rbrace_{n=1}^{N}$ is a 1D series of random variables with realizations $\lbrace t_n\rbrace_{n=1}^{N}$ of timestamp values corresponding to photon detection events. Eq. \ref{eq: Poisson model} is defined as the Poisson point-process arrival-time model, or ``Poisson noise model.''

This idealized, nonhomogeneous Poisson photon-counting model is extended to account for more realistic detector behavior, particularly nonextended deadtime. Extended deadtime is not discussed in this paper, but prior work has been done to model the behavior \cite{Muller}. Assuming the detector is active at $t=0$, then the arrival-time probability distribution function (PDF) of $T_1$ is known to be
\begin{equation} \label{eq: first photon}
    P\left(T_1=t_1\right)=\lambda(t_1)\exp\left(-\int_{0}^{t_1}\lambda(t')dt'\right).
\end{equation}
The detector undergoes deadtime after the first detection event, so the probability of detection is nulled during the deadtime interval. The PDF for $T_2$ conditioned on the arrival time $T_1$ is
\begin{equation}
\begin{split}
    P\left(T_2=t_2|T_1=t_1\right)=&\textbf{1}(t_2-t_1-\tau)\lambda(t_2)\\
&\times\exp\left(-\int_{t_1+\tau}^{t_2}\lambda(t')dt'\right),
\end{split}
\end{equation}
where $\tau$ is the deadtime interval and $\textbf{1}(t)$ is the unit step function. For convenience,
\begin{equation} \label{eq: integral}
    \Lambda(t)\triangleq\int_{0}^{t}\lambda(t')dt'
\end{equation}
such that the joint probability of the arrival of the first two photons is
\begin{equation}
\begin{split}
    P\left(\lbrace T_n=t_n\rbrace_{n=1}^{2}\right)=&\lambda(t_1)\exp\left[-\Lambda(t_1)\right]\\
    &\times\textbf{1}(t_2-t_1-\tau)\lambda(t_2)\\
    &\times\exp\left(-\Lambda(t_2)+\Lambda(t_1+\tau)\right).
\end{split}
\end{equation}
This can be extrapolated to the joint probability for $N$ detections as
\begin{equation}\label{deadtime_noise_model}
\begin{split}
    P\left(\lbrace T_n=t_n\rbrace_{n=1}^{N}\right)= & \lambda(t_1)\exp\left[-\Lambda(t_1)\right]\\
& \times\prod_{n=2}^{N}\lbrace\textbf{1}(t_n-t_{n-1}-\tau)\lambda(t_n)\\
&\times\exp\left[-\Lambda(t_n)+\Lambda(t_{n-1}+\tau)\right]\rbrace.
\end{split}
\end{equation}
Eq. \ref{deadtime_noise_model} is defined as the ``deadtime noise model,'' which is a modification of Eq. \ref{eq: Poisson model} (the Poisson noise model) to include deadtime. By defining
\begin{equation} \label{total lambda}
    \Lambda_T\left(\boldsymbol{t}\right)\triangleq\Lambda(t_1)+\sum_{n=2}^{N}\left[\Lambda(t_n)-\Lambda(t_{n-1}+\tau)\right],
\end{equation}
where $\boldsymbol{t}\triangleq\lbrace T_n=t_n\rbrace_{n=1}^{N}$ is the timestamp sequence of detections, then the deadtime noise model can be written as 
\begin{equation} \label{NHPPP Deadtime}
    P\left(\boldsymbol{t}\right)=\exp\left[-\Lambda_T\left(\boldsymbol{t}\right)\right]\prod_{n=1}^{N}\lambda(t_n).
\end{equation}

\subsubsection{Deadtime Noise Model: Discrete Form}

With the continuous form of the deadtime model defined, the next step is to modify the formulation to its discrete form. This will be useful because photon-counting systems commonly employ histogram acquisition systems, which are discrete samplers. Eq. \ref{total lambda} is rewritten as
\begin{equation}
\begin{split}
    \Lambda_T(\boldsymbol{t})=&\int_{0}^{t_N}\lambda(t')\\
    &\times\left\{1+\sum_{n=1}^{N}\left[-\textbf{1}(t'-t_n)+\textbf{1}(t'-t_n-\tau)\right]\right\}dt'
\end{split}
\end{equation}
by leveraging the unit step function $\textbf{1}(t)$ and Eq. \ref{eq: integral}. Substituting this back into Eq. \ref{NHPPP Deadtime} yields
\begin{equation} \label{2D deadtime}
\begin{split}
    &P\left(\boldsymbol{t}\right)= \prod_{n=1}^N\lambda(t_n) \exp\Biggl(-\int_{0}^{t_N}\lambda(t')\\
    &\times\left\{1+\sum_{n=1}^{N}\left[-\textbf{1}(t'-t_n)+\textbf{1}(t'-t_n-\tau)\right]\right\}dt'\Biggr).
\end{split}
\end{equation}
This is now modeled as a digitized process using the zero-order hold assumption (typical for digitization), where the arrival rate $\lambda(t)$ is constant within the bin interval $[t_m, t_{m+1})$ such that $\lambda(t_m)=\lambda_m$: 
\begin{equation} \label{eq: deadtime model}
    P\left(\boldsymbol{t}\right)=\exp\left(-\sum_{m=1}^{M}\lambda_{m}z_{m}\Delta t\right)\prod_{n=1}^{N}\lambda(t_{n}),
\end{equation}
where
\begin{equation} \label{eq: deadtime hist}
\begin{split}
    z_{m}\triangleq &\frac{1}{\Delta t}\int_{t_m}^{t_{m+1}}\biggl\{1+\sum_{n=1}^{N}[-\textbf{1}(t'-t_{n})\\
    &+\textbf{1}(t'-t_{n}-\tau)]\biggr\}dt'.
\end{split}
\end{equation}
$z_m$ is defined as the ``single-shot active histogram,'' where $z_m\in[0,1]$. The values at each bin indicate the state of the bin after a single laser shot, e.g., $z_m=0$ means that the detector was inactive during bin $m$ (due to deadtime from a previous detection), and $z_m=1$ means that the detector was active during bin $m$. Under the zero-order hold assumption, the remaining continuous term can be discretized as
\begin{equation}
    \prod_{n=1}^{N}\lambda(t_n)=\exp\left(\sum_{m=1}^{M}y_m\ln\lambda_m\right),
\end{equation}
where $y_{m}$ is defined as
\begin{equation} \label{eq: detection hist}
    y_{m}\triangleq\int_{t_m}^{t_{m+1}}\sum_{n=1}^{N}\delta(t'-t_n)dt',
\end{equation}
which is recognized as the standard photon-counting histogram. This will be a helpful formulation when deriving the histogram model in Sec. \ref{subsubsec: histogram MLE}, where $y_m$ is an available data product from histogram data-acquisition systems, i.e., multi-channel scalars. This results in the complete discrete form of the deadtime noise model:
\begin{equation} \label{eq: deadtime model}
    P\left(\boldsymbol{t}\right)=\exp\left[-\sum_{m=1}^{M}\left(\lambda_{m}z_{m}\Delta t-y_m\ln\lambda_m\right)\right].
\end{equation}
Note that this digitized formulation makes M\"uller Approximation \#1 by assuming a uniform flux in each bin when applying the zero-order hold approximation. 

\subsection{Maximum-Likelihood Estimation with Deadtime} \label{subsec: deadtime MLE}

\subsubsection{Maximum-Likelihood Estimation: Single-shot Model}

To generate a maximum-likelihood estimator for deadtime-corrected photon flux, the likelihood function is derived using Eq. \ref{eq: deadtime model} as mathematically equivalent to the PDF but parameterized using the arrival rate $\boldsymbol{\lambda}$ instead of the timestamps $\boldsymbol{t}$. The loss function is simplified using the negative-log likelihood form: 
\begin{equation} \label{eq: 1D loss deadtime}
\begin{split}
    \mathcal{L}_D\left(\boldsymbol{\lambda}; \boldsymbol{t}\right)&=-\ln\left[P\left(\boldsymbol{t};\boldsymbol{\lambda}  \right)\right]\\
    &=\sum_{m=1}^{M}\left(\lambda_mz_{m}\Delta t-y_{m}\ln\lambda_m\right),
\end{split}
\end{equation}
where $\boldsymbol{\lambda}$ is the discrete arrival-rate function as a function of bin $m$, and $\mathcal{L}_D$ is the loss function for the discrete deadtime noise model, or ``deadtime loss function.'' This study will test the deadtime noise model in scenarios where deadtime bias is negligible, commonly encountered in atmospheric lidar when backscatter densities are low. As a result, it also would be helpful to test the loss function derived from the Poisson noise model (Eq. \ref{eq: Poisson model}, where deadtime is excluded). This can be derived by setting $\tau=0$ such that the single-shot active histogram (Eq. \ref{eq: deadtime hist}) reduces to $z_m=1$ $\forall$ $m$, resulting in the ``Poisson loss function'':
\begin{equation} \label{eq: 1D loss poisson}
    \mathcal{L}_P\left(\boldsymbol{\lambda}; \boldsymbol{t}\right)=\sum_{m=1}^{M}\left(\lambda_m\Delta t-y_{m}\ln\lambda_m\right).
\end{equation}

\subsubsection{Maximum-Likelihood Estimation: Histogram Model} \label{subsubsec: histogram MLE}

\begin{figure}[htp]
    \centering    \includegraphics[width=0.45\textwidth]{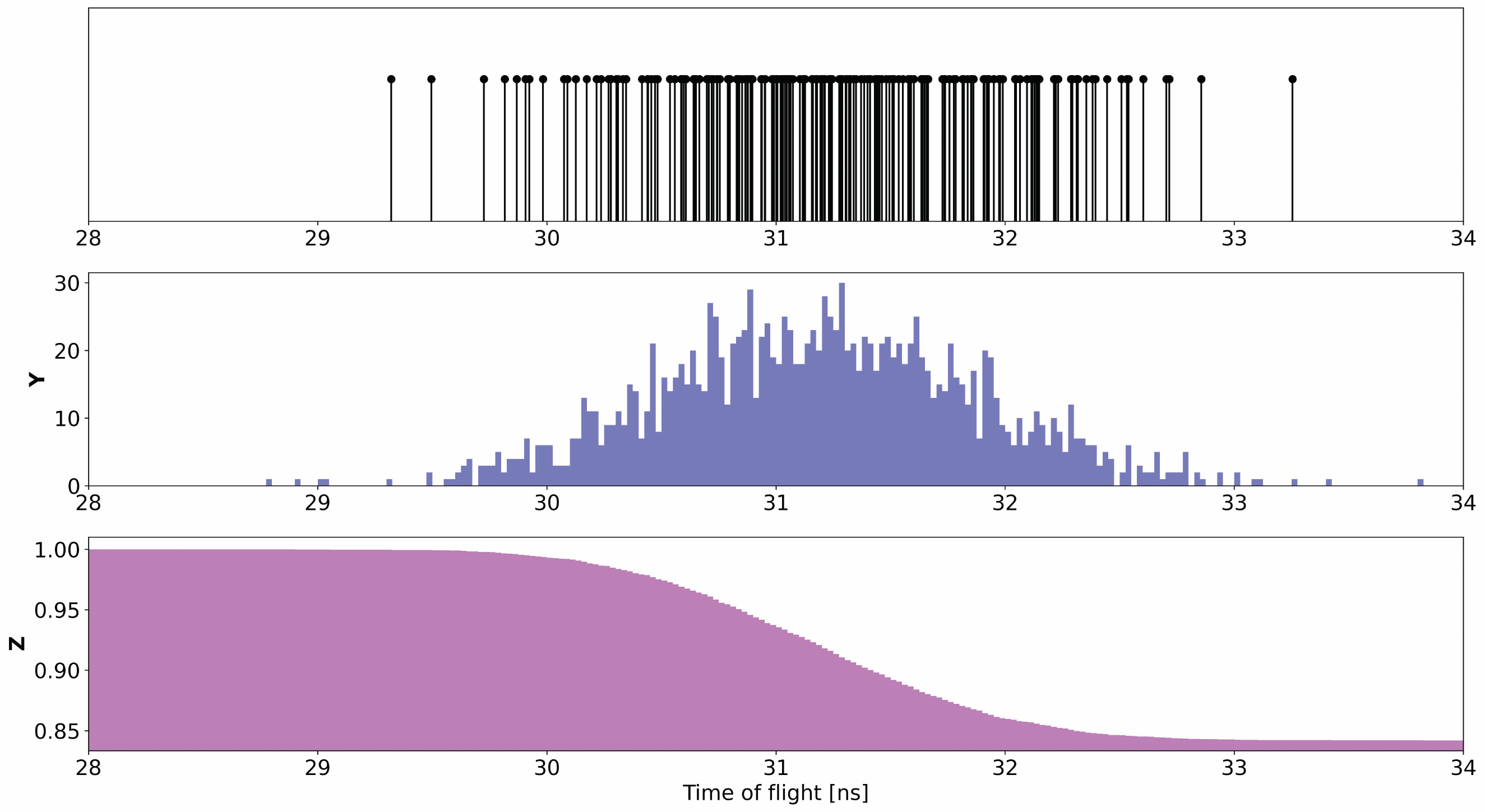}
    \caption{(Top) Photon counts: First 200 detected photons of a simulated Gaussian target (with a deadtime interval of 29 ns). (Center) $\textbf{Y}$ (Standard photon counting histogram): The sum of total photon counts per bin over the accumulation interval. (Bottom) $\textbf{Z}$ (active-fraction histogram): The average time the detector was active per bin over the accumulation interval. The value starts at unity, where there are sparse counts, and drops when the count density increases, a consequence of deadtime occupying bins that follow the detection bin.}
    \label{fig: Z_m Y_m}
\end{figure}

Extending the model to a histogram formulation is useful because photon-counting data is typically generated by accumulating signals over multiple laser shots and storing the aggregate in a histogram. Because the flux from each shot is independent, the loss function can be extended by summing the 1D deadtime loss function (Eq. \ref{eq: 1D loss deadtime}) over $\mathcal{N}$ laser shots:
\begin{equation}
    \mathcal{L}_D\left(\boldsymbol{\lambda}; \boldsymbol{t'}\right)=\sum_{i=1}^{\mathcal{N}}\sum_{m=1}^{M}\left(\lambda_{m,i}z_{m,i}\Delta t-y_{m,i}\ln\lambda_{m,i}\right),
\end{equation}
where $\boldsymbol{t'}\triangleq\left\lbrace\lbrace T_n=t'_n\rbrace _{n=1}^{N_i}\right\rbrace_{i=1}^{\mathcal{N}}$ and $t'_n=t_n-t_i$ is the relative timestamp to the most recent laser-shot timestamp $t_i$. This expression can be rewritten
\begin{equation} \label{eq: loss deadtime}
    \mathcal{L}_D\left(\boldsymbol{\lambda}; \boldsymbol{t'}\right)=\sum_{m=1}^{M}\left(\mathcal{N}\lambda_mZ_m\Delta t-Y_m\ln\lambda_m\right),
\end{equation}
where
\begin{equation} \label{eq: active-fraction histogram}
    Z_m\triangleq\frac{1}{\mathcal{N}}\sum_{i=1}^{\mathcal{N}}z_{m,i},
\end{equation}
\begin{equation}
    Y_m\triangleq\sum_{i=1}^{\mathcal{N}}y_{m,i},
\end{equation}
and the flux is assumed to be constant between shots $\lambda_{m,i}=\lambda_m$. $Z_m$ is an element of the normalized histogram $\textbf{Z}$ (Eq. \ref{eq: active-fraction histogram}), defined as the ``active-fraction histogram.'' It is a vector containing values for each bin of the single-shot active histogram (Eq. \ref{eq: deadtime hist}), averaged over $\mathcal{N}$ laser shots. Simply put, the value for each bin $m$ indicates what fraction the detector was active during that bin over the accumulation interval. $Y_m$ is an element of $\textbf{Y}$, recognized as the standard photon-counting histogram. An example of these histograms is visualized in Fig. \ref{fig: Z_m Y_m}. Finally, the estimated flux $\boldsymbol{\tilde{\lambda}}$ can be found by minimizing $\mathcal{L}_D$:
\begin{equation}
    \boldsymbol{\tilde{\lambda}}=\argmin_{\boldsymbol{\lambda}}\mathcal{L}_D\left(\boldsymbol{\lambda}; \boldsymbol{t'}\right).
\end{equation}
Eq. \ref{eq: loss deadtime} can be extended for the Poisson loss function (Eq. \ref{eq: 1D loss poisson}) as well:
\begin{equation} \label{eq: loss poisson}
    \mathcal{L}_P\left(\boldsymbol{\lambda}; \boldsymbol{t'}\right)=\sum_{m=1}^{M}\left(\mathcal{N}\lambda_m\Delta t-Y_m\ln\lambda_m\right).
\end{equation}

\subsection{Consistency with the M\"uller Correction} \label{subsec: Muller correction}

Under conditions where both M\"uller Approximations are satisfied (i.e., the observed signal is homogeneous over the sampling interval \textit{and} the bin width is much longer than the deadtime interval), then the deadtime noise model reduces to the M\"uller Correction (Eq. \ref{eq: Muller correction}). To demonstrate this consistency, first by satisfying Approximation \#2 ($\Delta t\gg\tau$), it can be assumed that the active fraction in a bin is only altered by photon arrivals within that bin such that $Z_m$ (Eq. \ref{eq: active-fraction histogram}) reduces to the following:
\begin{equation}
    Z_{m} = 1-Y_{m}\frac{\tau}{\mathcal{N}\Delta t}.
\end{equation}
Substituting this definition and analytically minimizing  Eq. \ref{eq: loss deadtime} for $\lambda_m$, the following result is obtained: 
\begin{equation}
\begin{split}
    \tilde{\lambda}_m&=\argmin_{\lambda_m}\mathcal{L}_D\left(\lambda_m; \boldsymbol{t'}\right)\\
    &=\frac{Y_{m}/\Delta t}{\mathcal{N}-Y_{m}\tau/\Delta t}=\frac{R_m}{1-R_m\tau},
\end{split}
\end{equation}
where $\tilde{\lambda}_m$ is the optimal flux estimate, and $R_m\triangleq Y_m/(\mathcal{N}\Delta t)$ is the measured count rate for bin $m$. The M\"uller Correction is reproduced, demonstrating consistency between both corrections and showing that the M\"uller Correction can be interpreted as the analytic form of a maximum-likelihood estimator for a homogeneous count rate accumulated over sampling intervals much longer than the deadtime interval, thereby satisfying the M\"uller Approximations. Thus, the two correction techniques are consistent.

\section{Methods}

\subsection{Approach} \label{subsec: experiment design}

After deriving the noise model, it was essential to test its accuracy across a dynamic range of fluxes typically encountered in the atmosphere by a backscatter lidar. The deadtime noise model was subjected to controlled experiments where its performance could be baselined against the current state of the art and confidently attribute the resulting improvement to correction of deadtime effects. 

To accomplish this, a stationary target was illuminated and the receiver incrementally attenuated with a neutral density (ND) filter to obtain a discrete range of fluxes over multiple orders of magnitude. To evaluate the corrected signal performance, a low-flux (without deadtime bias) measurement of the same target was used for comparison, discussed in Sec. \ref{subsubsec: eval}. This design concept was simple and enabled the observation of noise-model behavior as a function of flux, which is described in the next section in more detail. The experiment design parameters are described in detail in Sec. \ref{subsec: experiment setup}.

Most of the performance evaluation in this work focuses on estimating fine scale features in a very short target which presents an inherent challenge to both Approximations \#1 and \#2.  However, to further demonstrate the deadtime noise model's applicability to atmospheric lidar, it was also demonstrated on an extended target in Sec. \ref{subsubsec: extended signals}.

\subsection{Processing Routine} \label{subsec: fit routine}

\subsubsection{Fitting Procedure} \label{subsubsec: fitting procedure}

Using Eq. \ref{eq: loss deadtime}, the measurements are processed to generate fits that correct for deadtime. This approach assumes no prior knowledge of the functional form of the signal (as is the case for most atmospheric signals). One approach for parametric estimation is to use a $J$-degree polynomial basis such that the photon flux $\lambda$ is modeled
\begin{equation} \label{eq: poly basis}
    \lambda(\textbf{c}_J, b, t)=\exp\left[\sum_{j=0}^Jc_j\mathcal{T}_j(t)\right] + b,
\end{equation}
where $b$ is the constant background amplitude, $\textbf{c}_J=\lbrace c_j\rbrace_{j=0}^{J}$ is the set of fit coefficients, and $\lbrace \mathcal{T}_j(t)\rbrace _{j=0}^{J}$ is the Chebyshev polynomial set:
\begin{equation}
    \mathcal{T}_0(t)=1,
\end{equation}
\begin{equation}
    \mathcal{T}_1(t)=t,
\end{equation}
\begin{equation}
    \mathcal{T}_{j+1}(t)=2t\mathcal{T}_j(t)-\mathcal{T}_{j-1}(t).
\end{equation}
Maximum-likelihood estimation (based on the assumed noise model) was employed to estimate the polynomial coefficients and background amplitude:
\begin{equation}
\tilde{\textbf{c}}_J,\tilde{b} = \argmin_{\textbf{c}_J,b} \mathcal{L}\left(\lambda_J(\textbf{c}_J,b,t); \textbf{t}^{(f)} \right),
\end{equation}
where the loss function $\mathcal{L}$ is given by Eq. \ref{eq: loss deadtime} for the deadtime noise model ($\mathcal{L}_D$) or Eq. \ref{eq: loss poisson} for the Poisson noise model ($\mathcal{L}_P$), $\textbf{t}^{(f)}$ is the timestamps of the fit dataset, and $\tilde{\textbf{c}}_J$ and $\tilde{b}$ are the optimal fitting parameters for the specific polynomial basis of order $J$. 

% For the experiments in this study, the magnitude of the background signal was negligible compared to the signal of interest to ensure that the optimization process would be convex. 

Note that the polynomial order $J$ is an optimization variable too because a priori knowledge of the functional form of the signal is absent in most atmospheric-lidar applications. The routine employed holdout cross-validation to select the optimal polynomial order without overfitting, as in Ref. \cite{Hayman-holdout}. This was accomplished by evaluating the fit solution against a validation dataset (an independent, statistically identical dataset), which would produce a quantitative validation loss for a unique polynomial order $J$. Manual thinning was employed to generate the fit and validation datasets, where the time-tag data from alternating laser shots were assigned to the fit and validation datasets \cite{Hayman-2D-estimation}. The statistical properties of the fit and validation sets were equivalent, as a short interpulse period was used to oversample the stationary target. The optimal polynomial order $\tilde{J}$ would be selected based on the corresponding model with the minimum validation loss:
\begin{equation}
\tilde{J} = \argmin_{J} \mathcal{L}\left(\lambda(\tilde{\textbf{c}}_J,\tilde{b},t);\textbf{t}^{(v)}\right),
\end{equation}
where $\textbf{t}^{(v)}$ are the timestamps of the validation set. Finally, the optimal estimate $\tilde{\lambda}(t)$ was selected based on this optimal polynomial order:
\begin{equation}
    \tilde{\lambda}(t)\triangleq\lambda(\tilde{\textbf{c}}_{\tilde{J}},\tilde{b},t).
\end{equation}
\subsubsection{Noise Model Evaluation} \label{subsubsec: eval}

An unbiased measurement was needed for a baseline comparison to quantify the error of an optimal estimate. Without access to an experimental truth, the experiment leveraged that photon counts from the detectors have a linear, unbiased relationship to the incident photon flux in regimes where the flux is much less than $1/\tau$. Using a stationary target during the experiments enabled the validation of an estimate of the object in a high-flux regime against the same (significantly attenuated) object at low flux, essentially measuring the object with negligible deadtime bias. This low-flux dataset will be referred to as the ``evaluation dataset.'' 

The evaluation dataset was generated by exposing the detector to a very low flux over a long integration period to suppress shot noise. Because the fit and evaluation datasets were captured at different attenuation levels, it was necessary to compensate for this difference by scaling the flux estimate by a constant. Theoretically, the scaling factor would be the inverse of the transmission of the receiver chain, but, in practice, this value was not known to high enough accuracy. Thus, without accurate knowledge of this scaling factor a priori, the optimal scaling factor $T_r^*$ was used, which was obtained by minimizing the Poisson loss function (derivation provided in Appendix \ref{subsec: eval loss scale factor}):
\begin{equation}
    T_r^{*} = \frac{1}{\mathcal{N}\Delta t}\frac{\sum_{m=1}^MY_{m}^{(e)}}{\sum_{m=1}^M\tilde{\lambda}_m},
\end{equation}
where $Y_{m}^{(e)}$ is the evaluation histogram generated from the evaluation dataset $\textbf{t}^{(e)}$, and $\boldsymbol{\tilde\lambda}$ is the original forward model being evaluated. The linearly scaled forward model for comparison with the evaluation data is obtained:
\begin{equation}
    \boldsymbol{\tilde{\lambda}}^*=T_r^*\boldsymbol{\tilde{\lambda}},
\end{equation}
where $\boldsymbol{\tilde{\lambda}}^*$ is the scaled forward model. The Poisson loss function $\mathcal{L}_P$ in Eq. \ref{eq: loss poisson} was used to evaluate the model against the observed evaluation dataset:
\begin{equation} \label{eq: eval loss}
    \mathcal{L}_{eval} \triangleq \mathcal{L}_P\left(\boldsymbol{\tilde{\lambda}}^*; \textbf{t}^{(e)} \right),
\end{equation}
where $\mathcal{L}_{eval}$ will be called the model's "evaluation loss." The Poisson loss function was used instead of the deadtime loss function because the evaluation dataset was composed of a measurement without deadtime bias, which the Poisson loss function reflects. As previously stated, comparing the fit to the low-flux, unbiased evaluation dataset is the most effective approach to quantify the quality-of-fit because there is no access to the truth (to high enough accuracy) in the experiment. Thus, the evaluation loss will be used as the validation technique for the rest of this study.

Lastly, the methodology employed here represents an alternative to approaches where an analog detector channel is added to the receiver to sample a portion of the backscatter light. While the method can only be successful with a static scene, it avoids several practical complications imposed by the analog detection approach. For example, adding a second detector channel always introduces inconsistencies (e.g., different detection efficiencies and overlap functions) that must be calibrated to high accuracy to obtain accurate evaluation metrics. Also, the analog detector does not provide a reliable estimate of truth at low flux, which limits its dynamic range and, thus, its utility as a validation technique. While these issues are not insurmountable, the approach described in this section provided a simple and robust alternative requiring no additional hardware.

\subsection{Experiment Setup} \label{subsec: experiment setup}

\begin{figure}[t]
    \centering    \includegraphics[width=0.47\textwidth]{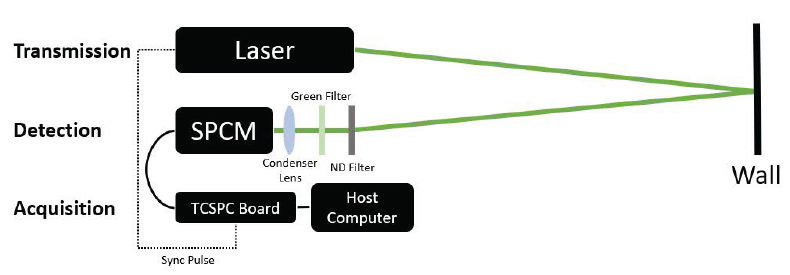}
    \caption{\textbf{Experiment Setup}: The pulsed laser illuminates the wall in which the diffuse reflection passes through the receiver chain composed of the ND filter, thin-film filter, and condenser lens, which focuses the remaining photons on the detection plane of the single-photon counting module (SPCM) detector. The output electrical signal from each detection is sent to the TCSPC acquisition electronics and stored on the host computer. The electrical laser sync pulse is also saved via the acquisition board. Signal types: Optical (green) and Electrical (black).}
    \label{fig: setup}
\end{figure}

Fig. \ref{fig: setup} shows the experiment design. This setup probed a static scene where the collected flux $\lambda_0(t)$ was consistent between measurements. Other variables (e.g., background light, temperature) were consistent between measurements. The wall was selected as the target because it was stationary between shots and did not affect the signal shape. It also increased the signal path length, which temporally separated the initial flash of the laser and the reflected beam, thus ensuring that the signal of interest was isolated from transmitter interference. Most importantly, the resultant backscatter signal would be very narrow in range ($<0.5$ m), thus containing the high spatial-frequency content needed to evaluate the noise model at high resolution.

To vary attenuation, an ND filter was inserted immediately in front of the receiver aperture so that the signal and background would be equally attenuated and the signal's shape would be preserved. According to Fig. \ref{fig: setup}, the time-of-flight dependent flux $\lambda_{obs}(t)$ incident on the detector face for a given ND filter is described by
\begin{equation} \label{eq: OD}
    \lambda_{obs}(t)=10^{-\mathrm{OD}}T_fT_l[\lambda_0(t)+\lambda_{b}]=T_{r}[\lambda_0(t)+\lambda_{b}],
\end{equation}
where $\mathrm{OD}$ is the optical density value of the ND filter, $T_f$ and $T_l$ are the transmissivities of the spectral filter and condenser lens, $\lambda_0(t)$ is the flux resulting from the laser light scattering from the target, $\lambda_{b}$ is the constant background flux, and $T_r$ is the total transmission of the receiver chain. $T_r$ was manually tuned by swapping or combining ND filters of different OD values, enabling the measurement of stationary targets of different amplitudes but identical pulse shapes.

The transmitted beam was directed toward the wall, which was four meters away, to ensure a sufficient time gap between the laser pulse exiting the instrument and the signal of interest and to avoid interference between the two events. The reflection from the wall was observed by the receiver optics composed of an ND filter, thin-film filter, and condenser lens to focus the light onto the detector plane. For the ND filter, the fit and evaluation datasets were generated using OD 1.8 - 3.3 (at 0.1 increments) and OD 5.0, respectively. The receiver system was assembled in a light-tight configuration to minimize transmitter- and background-light interference. The detector was the Excelitas Single-Photon Counting Module (SPCM). For each detection event, the SPCM would output an electrical transistor-transistor-logic (TTL) signal, which would be assigned a digital timestamp (relative to the start of acquisition) via PicoQuant TimeHarp 260 PICO TCSPC at 25-ps resolution. Because each laser shot was also assigned a time stamp, relative detection times were also generated. Like binning acquisition techniques, such as multi-channel scalar, histograms were constructed in post-processing using the native 25-ps resolution. The most relevant experiment specs are listed in Tab. \ref{tab: specs}.  Note that because the TCSPC acquisition system deadtime is less than the detector deadtime, only the detector deadtime needs to be be modeled in the detection process \cite{Rapp2021}.

\begin{table} \label{tab: experiment specs}
\centering
\caption{Most relevant experimental parameters, where $\lambda_c$ is the laser central wavelength and $f_L$ is the pulse repetition rate.}\label{tab: specs}    
\begin{tabularx}{0.47\textwidth}{X|>{\hsize=\hsize}X}
    \multicolumn{2}{c}{\textbf{Important Experiment Parameters}} \\
    \hline
    \textbf{Laser Transmitter} & $\lambda_c=532.18$ nm, $f_L=14.3$ kHz, Power $=35$ mW, FWHM$<700$ ps\\
    \textbf{SPCM Detector} & Pulse width $=17.1$ ns TTL, Deadtime $=29.1$ ns\\
    \textbf{TCSPC Acquisition} & Bin width = 25 ps, Deadtime $<25$ ns\\
    \end{tabularx}
\end{table}

\section{Experiments} \label{sec: Simulations}

\subsection{Simulations}

The experiments were first conducted in simulation to test the estimator to high precision and verify the experiment design, following the approach and experiment designs described in Sec. \ref{subsec: experiment design} and \ref{subsec: experiment setup}. The simulated transmitter pulse shape was a Gaussian signal with a full-width half-maximum (FWHM) value of $\sim$1.18 ns, which is shorter than the deadtime interval and narrow in range to represent features that are finer than those encountered in the atmosphere, as stated in Sec. \ref{subsec: experiment design}. The wall's reflection response function was reasonably approximated as a delta function, and the simulated signal amplitudes were based on flux amplitudes that closely matched those eventually observed in the lab. Detections were generated by simulating nonhomogeneous Poisson arrival times with deadtime using Cinlar's method \cite{Simulations-source}, which were acquired using 25-ps resolution (to match the experimental resolution) with a 25-ns non-extended detector deadtime. $10^6$ laser shots were simulated per histogram.

This work uses the active-fraction histogram $\mathbf{Z}$ (Eq. \ref{eq: active-fraction histogram}) as the variable defining dynamic range, which directly quantifies how often the deadtime interval from a previous bin disabled each bin. Thus, for the rest of this study, the independent variable used is the average fractional time within the measurement window that the detector was active (or ``active fraction''). Formally, the active fraction (AF) is defined as the mean value of the active-fraction histogram $\textbf{Z}$ from Sec. \ref{subsubsec: histogram MLE}:
\begin{equation}
    \mathrm{AF}\triangleq\langle \textbf{Z}\rangle=\frac{1}{M}\sum_{m=1}^MZ_{m},
\end{equation}
where $M$ is the total number of histogram bins. By keeping the measurement window length and bin width constant throughout this study (i.e., $M$ was constant), AF is a valuable metric for deadtime bias as it directly describes the deadtime effect on the observed photon counts. For example, deadtime infrequently deactivates the detector at low fluxes, resulting in AF $\approx1$. In contrast, deadtime frequently deactivates the detector at high fluxes, resulting in AF $<1$.

\begin{figure}[t!]
     \begin{subfigure}[t]{0.23\textwidth}
         \centering        \includegraphics[width=\textwidth]{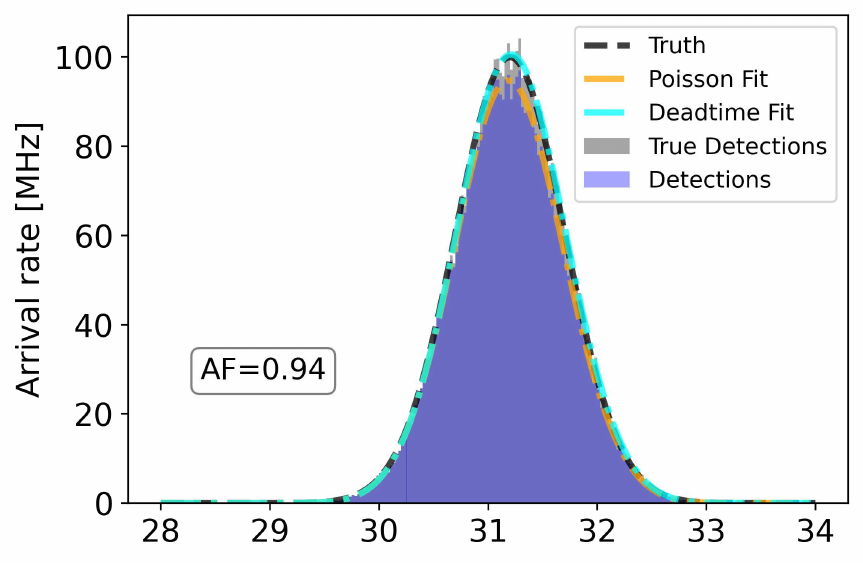}
         \caption{}
         \label{subfig: 100 MHz}
     \end{subfigure}
     \hfill
     \begin{subfigure}[t]{0.23\textwidth}
         \centering        \includegraphics[width=\textwidth]{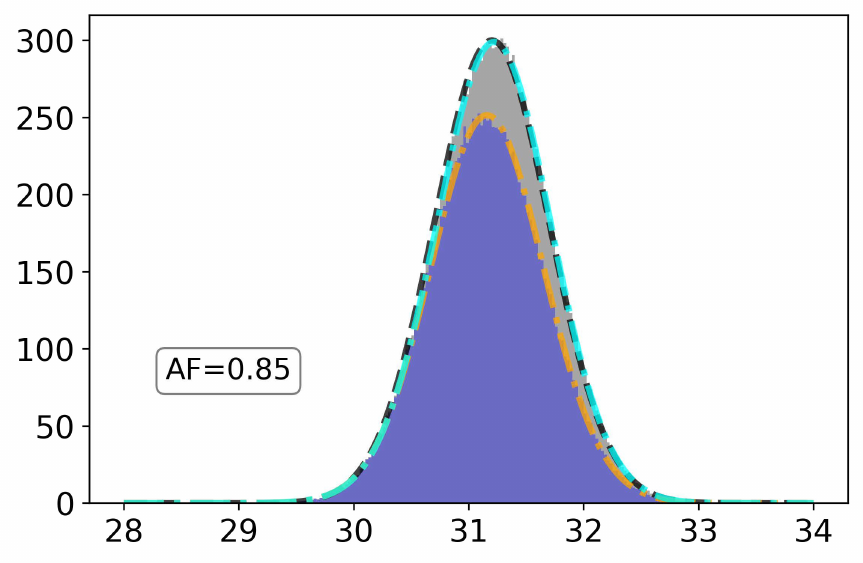}
         \caption{}
         \label{subfig: 300 MHz}
     \end{subfigure}
     \hfill
     \begin{subfigure}[t]{0.23\textwidth}
         \centering        \includegraphics[width=\textwidth]{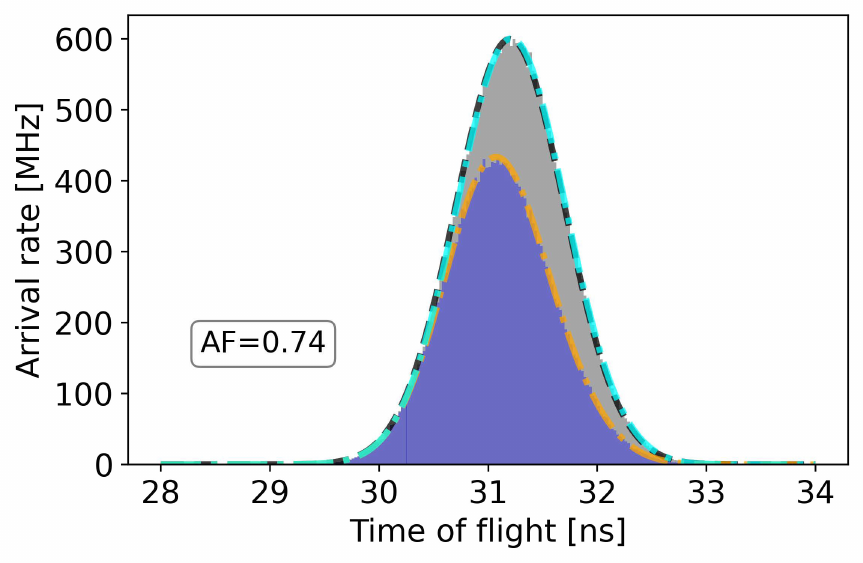}
         \caption{}
         \label{subfig: 600 MHz}
     \end{subfigure}
     \hfill
     \begin{subfigure}[t]{0.23\textwidth}
         \centering        \includegraphics[width=\textwidth]{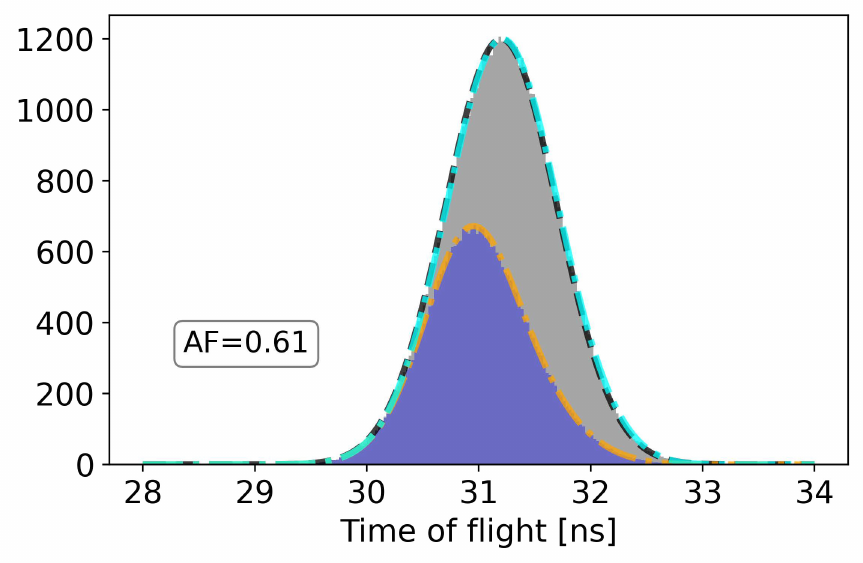}
         \caption{}
         \label{subfig: 1 GHz}
     \end{subfigure}
     
    \caption{These are simulated profiles with increasing peak arrival rates and the fits generated via the Poisson and Deadtime models. $10^6$ laser shots were used for each generated histogram. The corresponding AF values are also included, indicating a significant deadtime bias when AF $\lesssim0.85$.}
    \label{fig: simulated profiles}
\end{figure}

\subsubsection{Results}

Some example simulated profiles and their respective deadtime fits are shown in Fig. \ref{fig: simulated profiles}. The Poisson fits are also compared because they represent the standard estimation approach when the user assumes that deadtime bias is negligible, a valid assumption for applications in the low-flux regime \cite{Deadtime-Tomography}. In Fig. \ref{fig: simulated profiles}, the detector's corresponding AF value was included with each fit to provide context for how much each measurement was affected by deadtime (e.g., small AF indicates more considerable deadtime bias). 

At low flux (Fig. \ref{subfig: 100 MHz}), the simulated detection profile was approximately linear with respect to the true profile. Conversely, at higher flux rates (Fig. \ref{subfig: 300 MHz} - \ref{subfig: 1 GHz}), the detection profiles were biased low. Additionally, the locations of their peaks were shifted closer to the leading edge, a phenomenon known as first photon bias or range walk error \cite{FPB, Range-walk-error}. The deadtime fit corrected for these signal alterations, while they were left uncorrected by the Poisson fit. 

Though estimates from the M\"{u}ller Correction would be an intuitive candidate for comparison, they are omitted because the M\"{u}ller Correction produces non-physical, off-scale (often negative) photon-count estimates at these fine scales, due to violating Approximation \#2 (examples of this phenomenon are shown in Appendix \ref{subsec: Muller appendix}). Alternatively, if the bin width were dramatically increased to undersample the target, the M\"{u}ller Correction would still fail because this approach would violate Approximation \#1 (as in Fig. \ref{fig: Muller correction test}). In summary, the M\"{u}ller Correction simply fails to produce accurate estimates when violating the M\"{u}ller Approximations, as previously emphasized in Sec. \ref{subsec: Muller approx}, which is why the Poisson model is used for baseline comparison in the narrow-target studies. 

\subsubsection{Discussion} \label{subsubsec: sim discussion}

The results in Fig. \ref{fig: simulated profiles} confirm that the estimator generates accurate estimates of high-flux targets with high spatial-frequency content. This capability is unattainable when using the M\"{u}ller Correction. This exemplifies the folly of trying to correct for deadtime at resolutions too coarse to reflect the heterogeneity of the target. For example, undersampling the target with the highest peak flux ($\sim$650 MHz, Fig. \ref{subfig: 1 GHz}) by integrating the counts into a single 1-ns bin (instead of 25 ps in the figure) would produce a scalar mean flux of $\sim$800 kHz. For users processing at this coarse bin width, it would be challenging to recognize from this single value that the measurement is biased low by deadtime, thus naive application of the M\"{u}ller Correction to this singular measurement would be erroneous (due to the violation of Approximation \#1). 

\begin{figure}[h] \label{fig: RMSE vs eval loss}
     \centering
     \begin{subfigure}[h]{0.235\textwidth}
         \centering        
         \includegraphics[width=\textwidth]{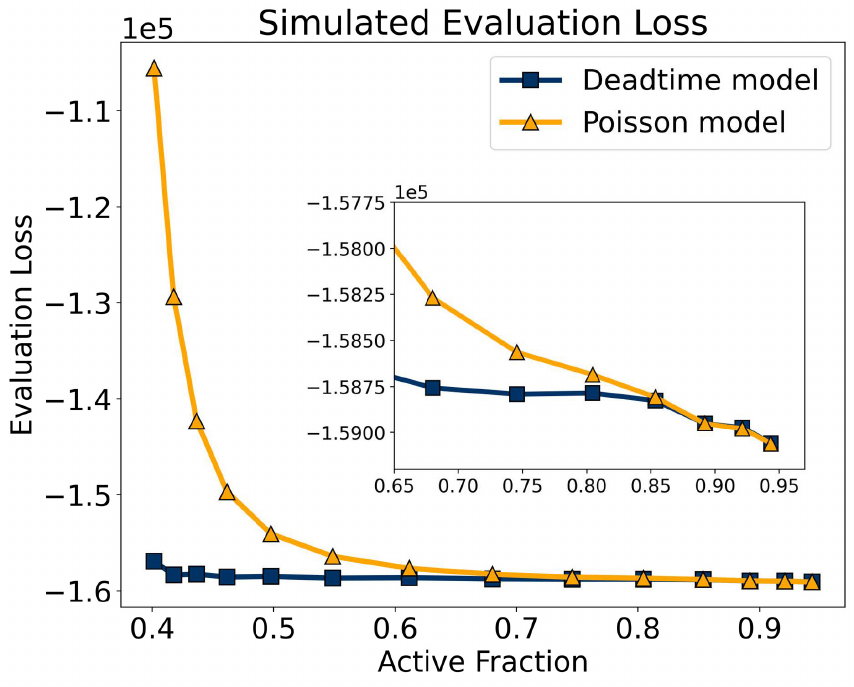}
         \caption{Evaluation Loss}
         \label{subfig: RMSE}
     \end{subfigure}
     \hfill
     \begin{subfigure}[h]{0.225\textwidth}
         \centering      \includegraphics[width=\textwidth]{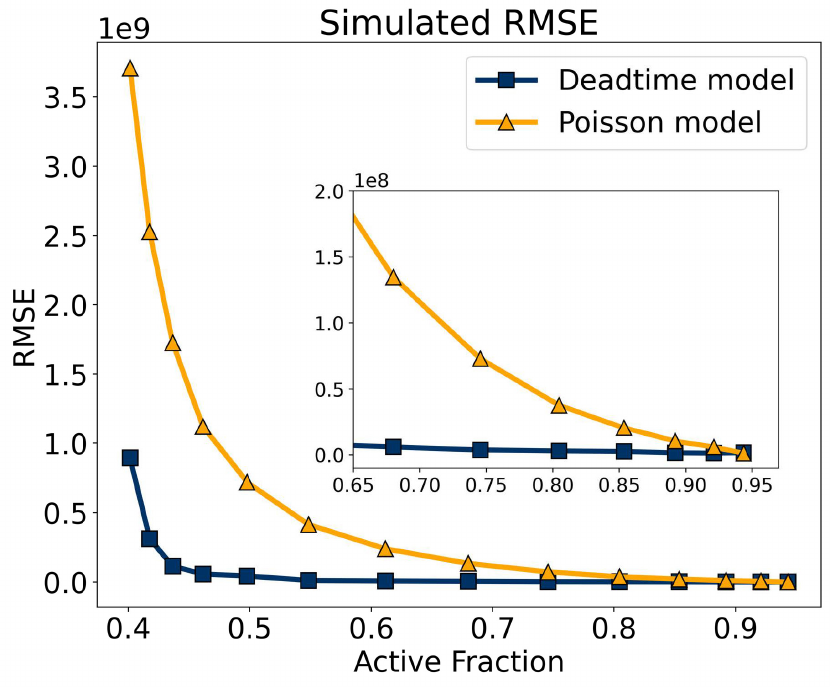}
         \caption{RMSE}
         \label{subfig: sim eval loss}
     \end{subfigure}
     \caption{Comparison of RMSE and evaluation loss values for the deadtime and Poisson fits across flux regimes. The number of laser shots used was $10^6$ for each fit.}
 \end{figure}

The deadtime and Poisson fitting routines' errors were calculated across incident fluxes using root-mean-square error (RMSE), shown in Fig. \ref{subfig: RMSE}. The most important result from the RMSE curve is that the error for the deadtime fit stays relatively constant across AF values, indicating high accuracy from low to high flux (or high to low AF, equivalently). Eventually, the error increases at very high fluxes, where the reason is discussed in more detail in Sec. \ref{subsubsec: exp discussion}. Meanwhile, the Poisson fit diverges rapidly at higher fluxes (or lower AF), which was expected since deadtime bias is non-negligible at high flux.

As established previously in Sec. \ref{subsec: experiment design}, the evaluation loss (obtained by comparing the estimate against a separate evaluation dataset) was selected as the accuracy metric for each fit since the truth is not known to high accuracy during actual measurements. The evaluation dataset was generated by simulating a signal with a 1-MHz peak arrival rate (low flux), accumulated over $10^7$ laser shots (an order of magnitude longer than the other measurements to mitigate shot noise). The evaluation losses for each fit were calculated and plotted in Fig. \ref{subfig: sim eval loss}. Like the RMSE curves, the deadtime fit's evaluation loss was relatively constant across a wide range of fluxes, while the Poisson fit's loss rapidly increased at higher fluxes. Also, the monotonic relationship between RMSE and the evaluation loss should be noted.

 \subsection{Measurements} \label{subsec: experiment results}

\begin{figure}[h]
     \centering
     \begin{subfigure}[h]{0.375\textwidth}
         \centering        
         \includegraphics[width=1\linewidth]{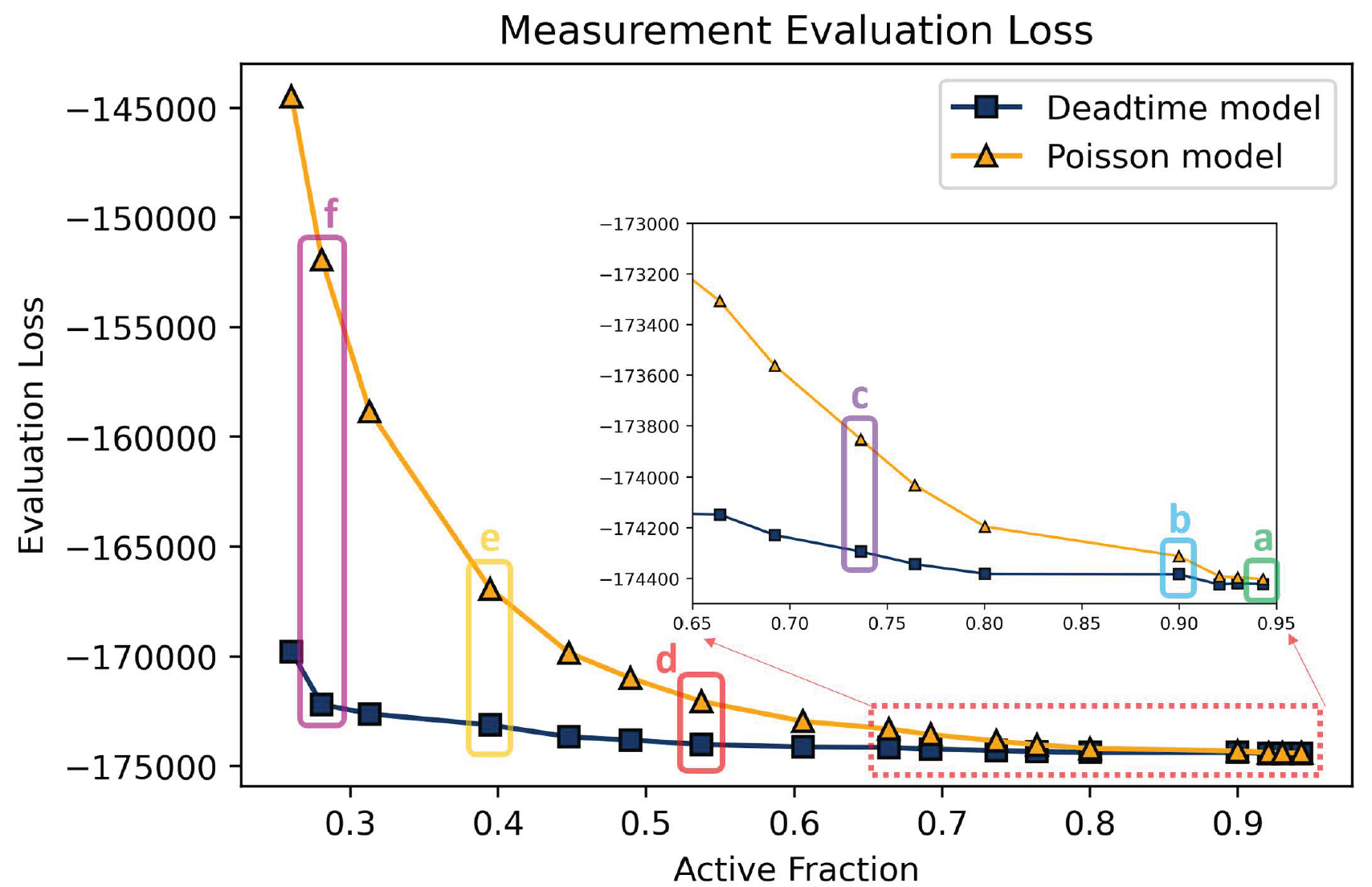}
     \end{subfigure}
     \hfill
     \begin{subfigure}[b]{0.475\textwidth}
         \centering      \includegraphics[width=1\linewidth]{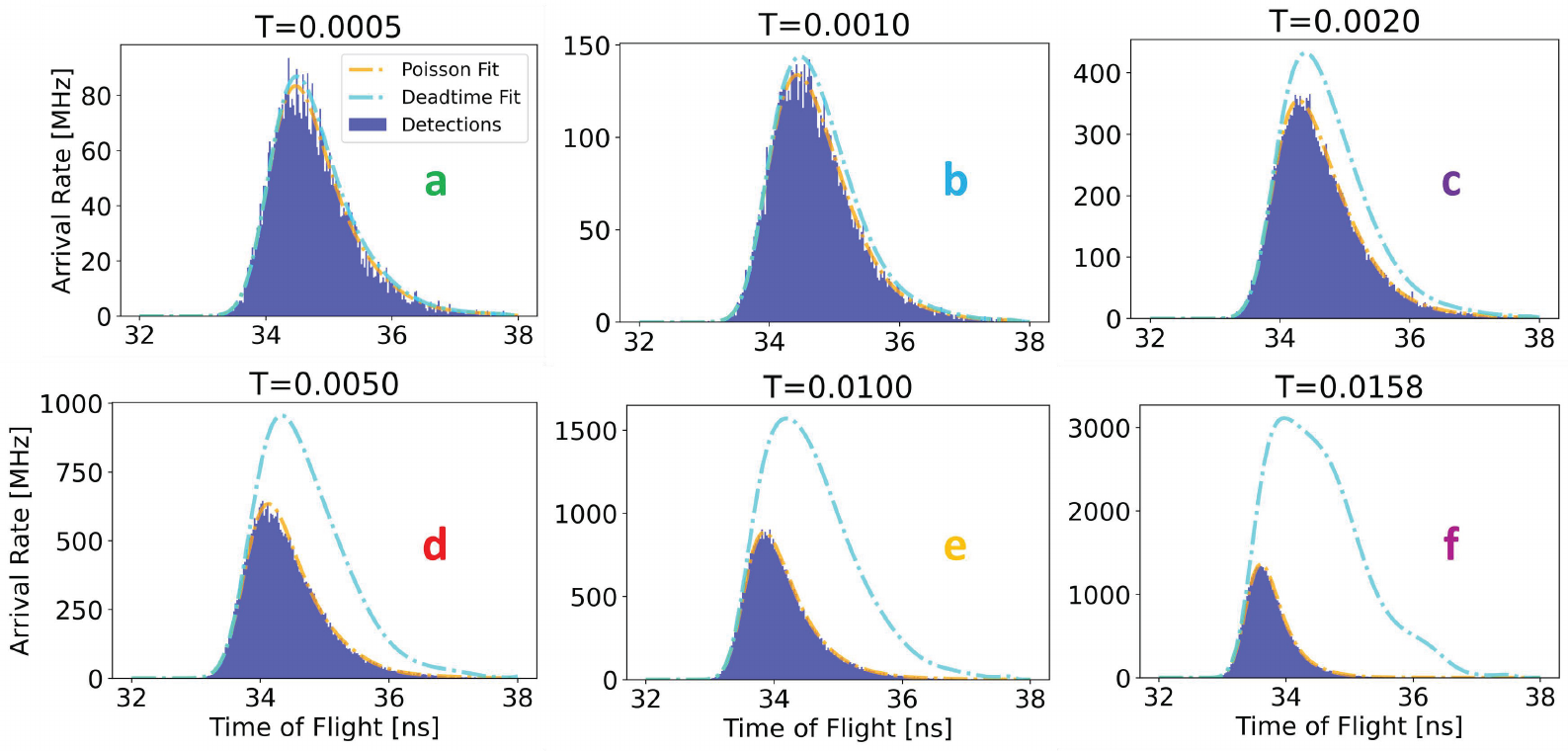}
     \end{subfigure}
     \caption{Evaluation losses for the deadtime and Poisson fits are included. Examples of the fits at different AF values and their respective transmission values are also included for reference. The transmission value $T$ is related to the OD of the ND filter for the measurement by $T=10^{-\mathrm{OD}}$. Each model's evaluation loss curve corresponds to fit quality. The deadtime fit compensates for the flux attenuation and first photon bias, both nonlinear features of deadtime bias.}
     \label{fig: histogram examples}
 \end{figure}

Two-minute measurements ($\sim1.7\times10^6$ laser shots) were collected for each attenuation value. For validation, the evaluation dataset (described in Sec. \ref{subsubsec: eval}) was obtained using a very low-flux measurement accumulated over $1.4\times10^7$ shots. Data was processed following that described in Sec. \ref{subsubsec: fitting procedure}. This was done for each attenuation value individually by calculating the optimal fit and then generating an evaluation-loss value for the fit by comparing the fit solution against an unbiased (low flux) evaluation dataset. These values are plotted in Fig. \ref{fig: histogram examples}. Recalling that a more negative evaluation loss indicates higher accuracy, Fig. \ref{fig: histogram examples} shows that the fits obtained using the deadtime model produced more accurate estimates of the actual profile than the Poisson model. 

 \subsubsection{Discussion} \label{subsubsec: exp discussion}

The experimental and simulated results demonstrated that it was incorrect to assume that deadtime bias was negligible for relatively low fluxes, resulting in a high active-detector time (or $\mathrm{AF}\geq90\%$). This was shown in Fig. \ref{fig: histogram examples}a-\ref{fig: histogram examples}c, where visually minor differences between the Poisson and deadtime evaluation loss curves corresponded to noticeable deadtime corrections in their respective fits. For example, when the detector was 90\% active on average ($\mathrm{AF}=90\%$ in Fig. \ref{fig: histogram examples}b), the deadtime and Poisson fits differed noticeably, indicating that deadtime bias was non-negligible even in a flux regime that could be considered qualitatively ``linear.'' This was also observed in the simulated cases where the deadtime fit still outperformed the Poisson fit even with $\mathrm{AF}=0.94$ (Fig. \ref{subfig: 100 MHz}). This outcome implies that due to the variety and variability of atmospheric targets, users without a viable deadtime correction method are forced to identify which measurements qualify as linear and which do not during data processing. This means that high-flux targets (e.g., dense aerosol layers, clouds) must be removed from the data to avoid inaccurate retrievals, ultimately sacrificing data availability. However, for users that leverage the deadtime estimator introduced in this study with high-capture resolution in range and time, data removal may not be necessary. Instead, the deadtime fit offers newfound flexibility where the user can process data from observations that span an extensive dynamic range without having to identify and remove biased data.

Also, in Fig. \ref{fig: histogram examples}f, the deadtime fit is qualitatively poor in the estimate's falling edge, indicating a limitation with the approach. This degradation is explained by the detector's saturation induced by high backscatter at the target's leading edge. The high rate at which deadtime was triggered from the leading edge's signal resulted in sparse to zero counts from the falling edge during accumulation. This meant the fit to the falling edge was noisy, which was caused by fitting to shot noise in this region. This indicates that the estimator performance suffers somewhat in regions where detector saturation results in high shot noise in subsequent bins. Because this is ultimately a shot-noise-induced error, it also suggests that the correction technique will be sensitive to other types of measurements that contain shot noise. One example is optically thick targets, whose rapid extinction can induce shot noise at large optical depths. Another example is insufficient accumulation intervals, which is discussed further in the next section.

Overall, these results highlight the promise of the deadtime model: when observing scenes with variable backscatter intensities, the deadtime-fit routine can enable accurate estimates at high resolution across a wide range of fluxes and spatial variabilities. This new capability shows promise to expand the utility of photon-counting systems across a more extensive array of atmospheric targets, especially those containing large spatial gradients in backscatter intensity. 

\subsubsection{Shot Noise Impact on Fit Quality} \label{subsec: shot limit}

As discussed in the previous section, the estimator's accuracy diminishes in the presence of shot noise. A common source of this in photon counting is incompatible bin widths (range or time) for the observed signal strength. Although the estimator can enable high-resolution retrievals, the potential downside of using small-range bin widths is that shot noise increases as bin width decreases for a fixed accumulation length. Strategies to mitigate shot noise typically involve increasing transmitted power, receiver aperture diameter, or expanding the integration length in time or range. However, these options can be infeasible or unattractive for many applications, especially those where the laser repetition rate is low or high temporal resolution is necessary, which are common scenarios in atmospheric lidar. Therefore, analyzing how the estimator performs on measurements containing shot noise is essential. As introduced previously, saturation on the detector (from the signal's leading edge, rapid extinction, insufficient accumulation intervals, etc.) can also induce shot noise.

\begin{figure}[h]
     \begin{subfigure}[t]{0.235\textwidth}
         \centering        \includegraphics[width=\linewidth]{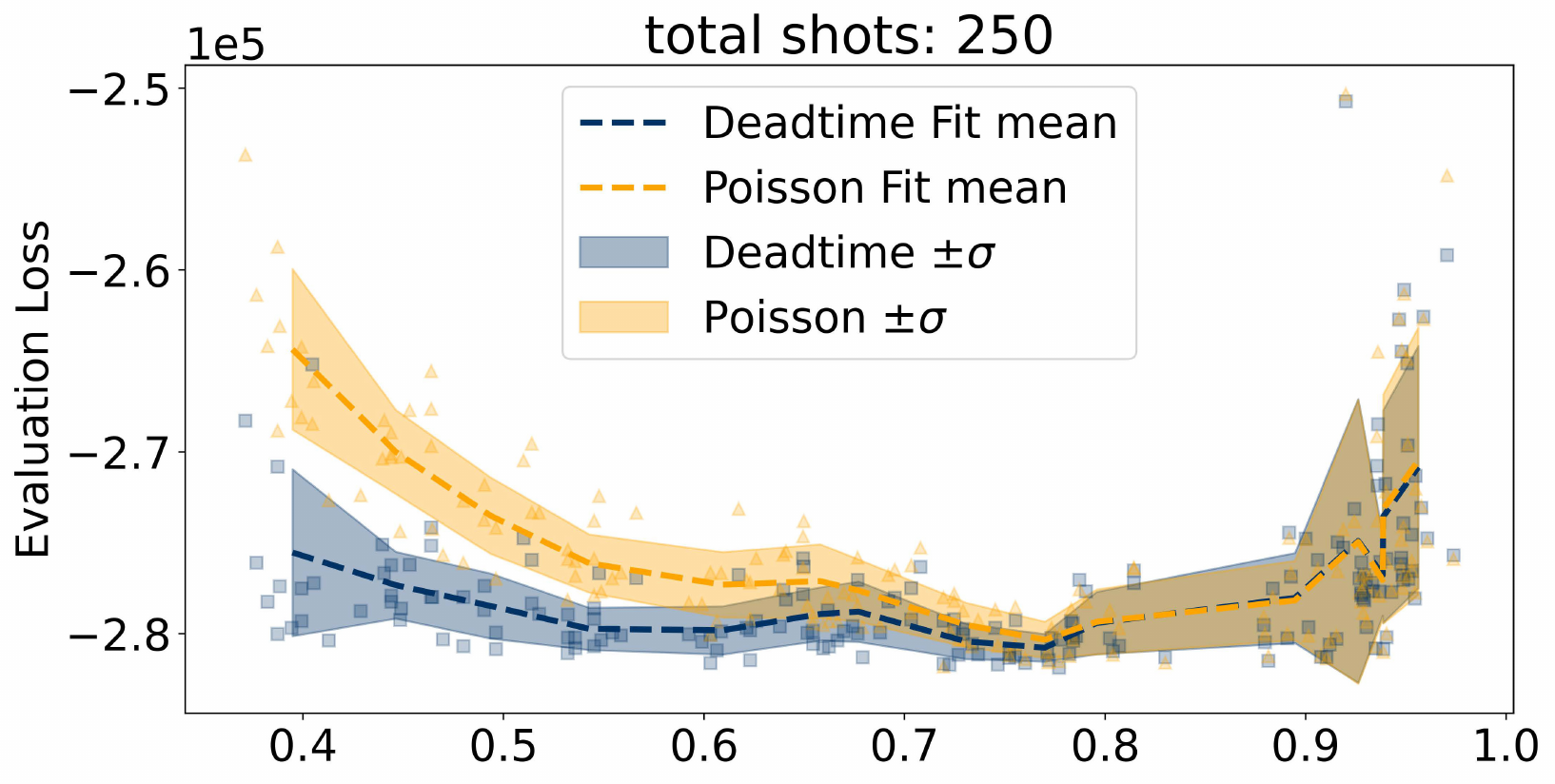}
         \label{subfig: spread 250}
     \end{subfigure}
     \hfill
     \begin{subfigure}[t]{0.225\textwidth}
         \centering        \includegraphics[width=\linewidth]{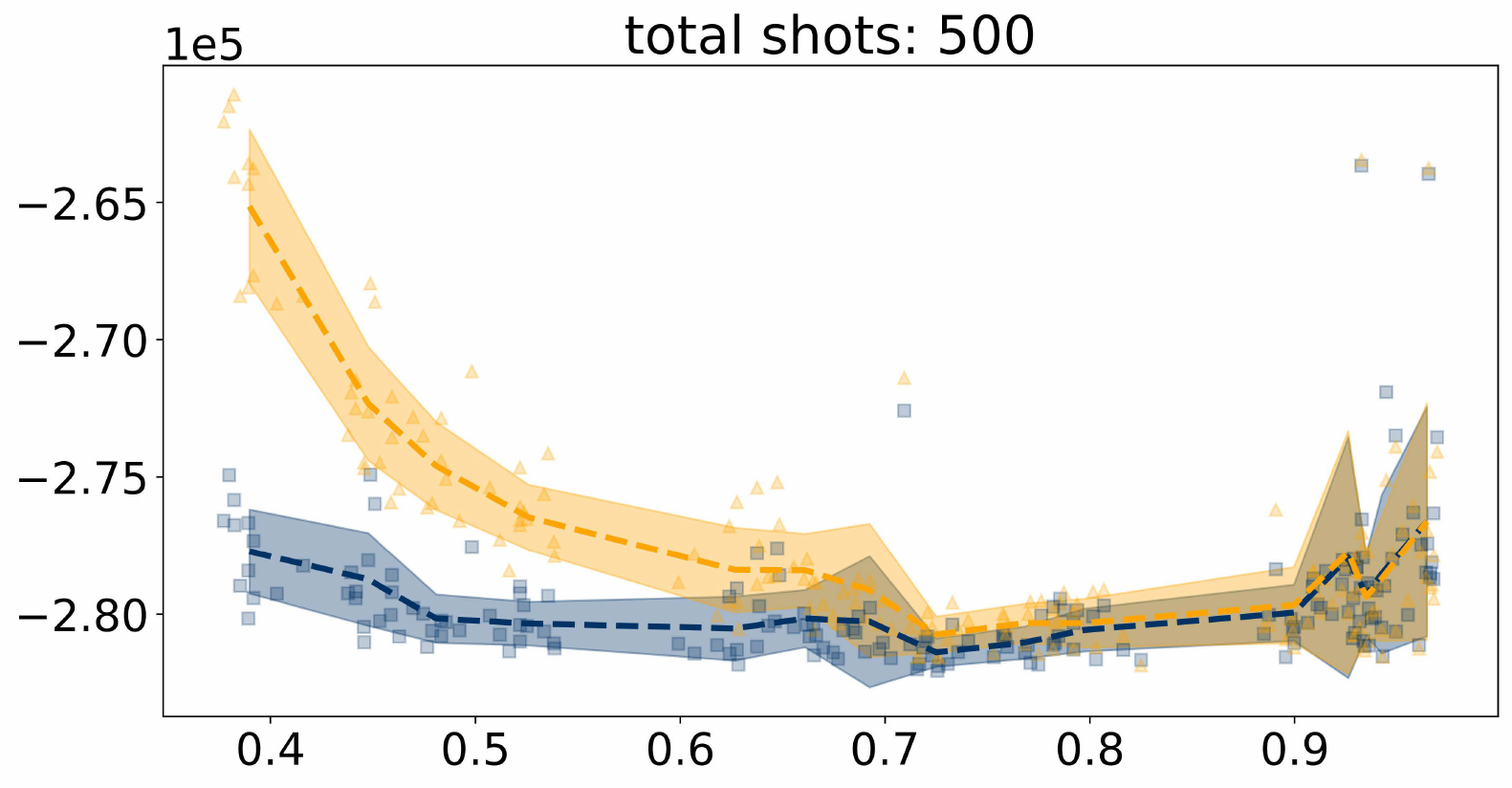}
         \label{subfig: spread 500}
     \end{subfigure}
     \hfill
     \begin{subfigure}[t]{0.235\textwidth}
         \centering        \includegraphics[width=\linewidth]{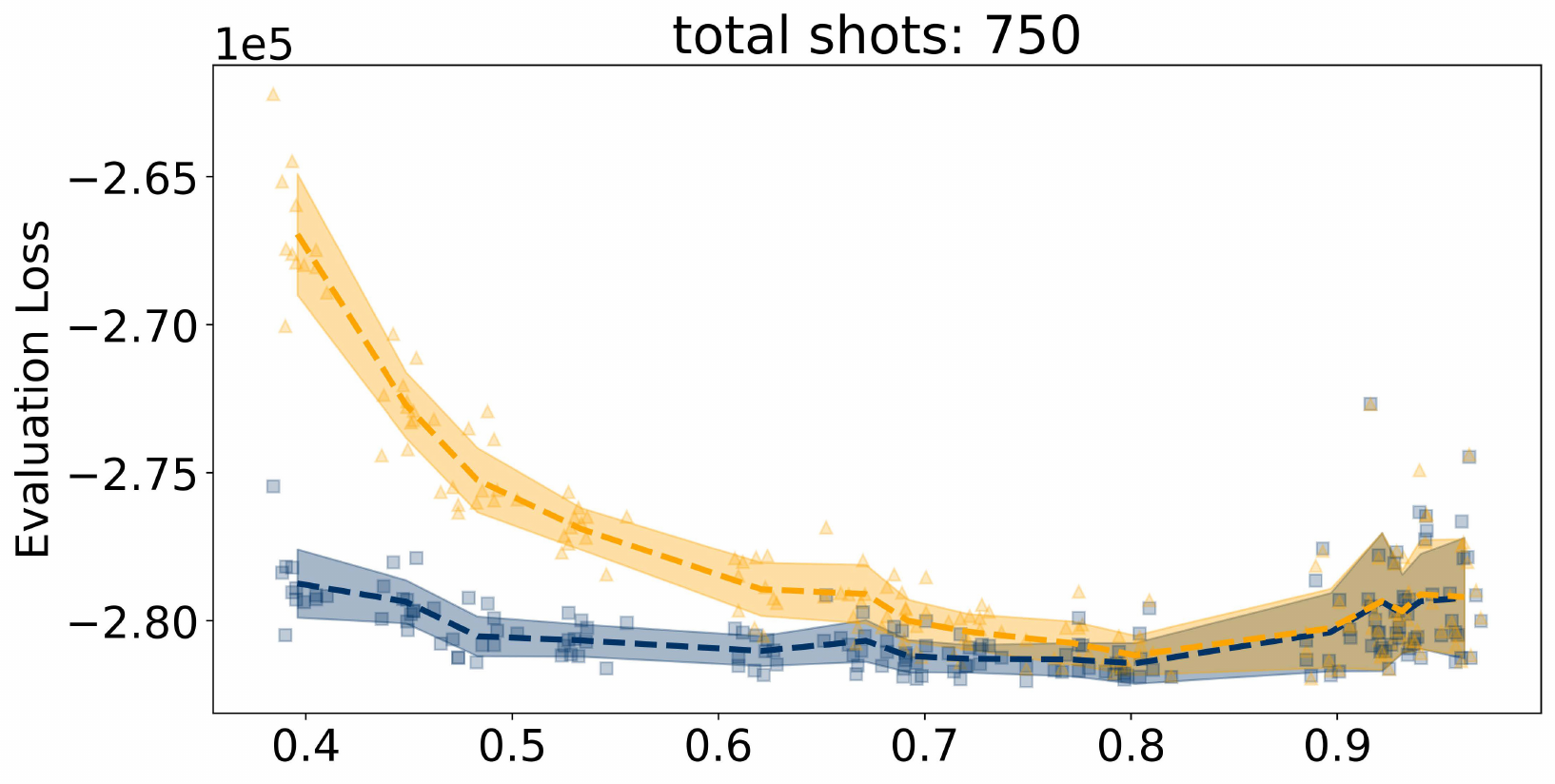}
         \label{subfig: spread 750}
     \end{subfigure}
     \hfill
     \begin{subfigure}[t]{0.225\textwidth}
         \centering        \includegraphics[width=\linewidth]{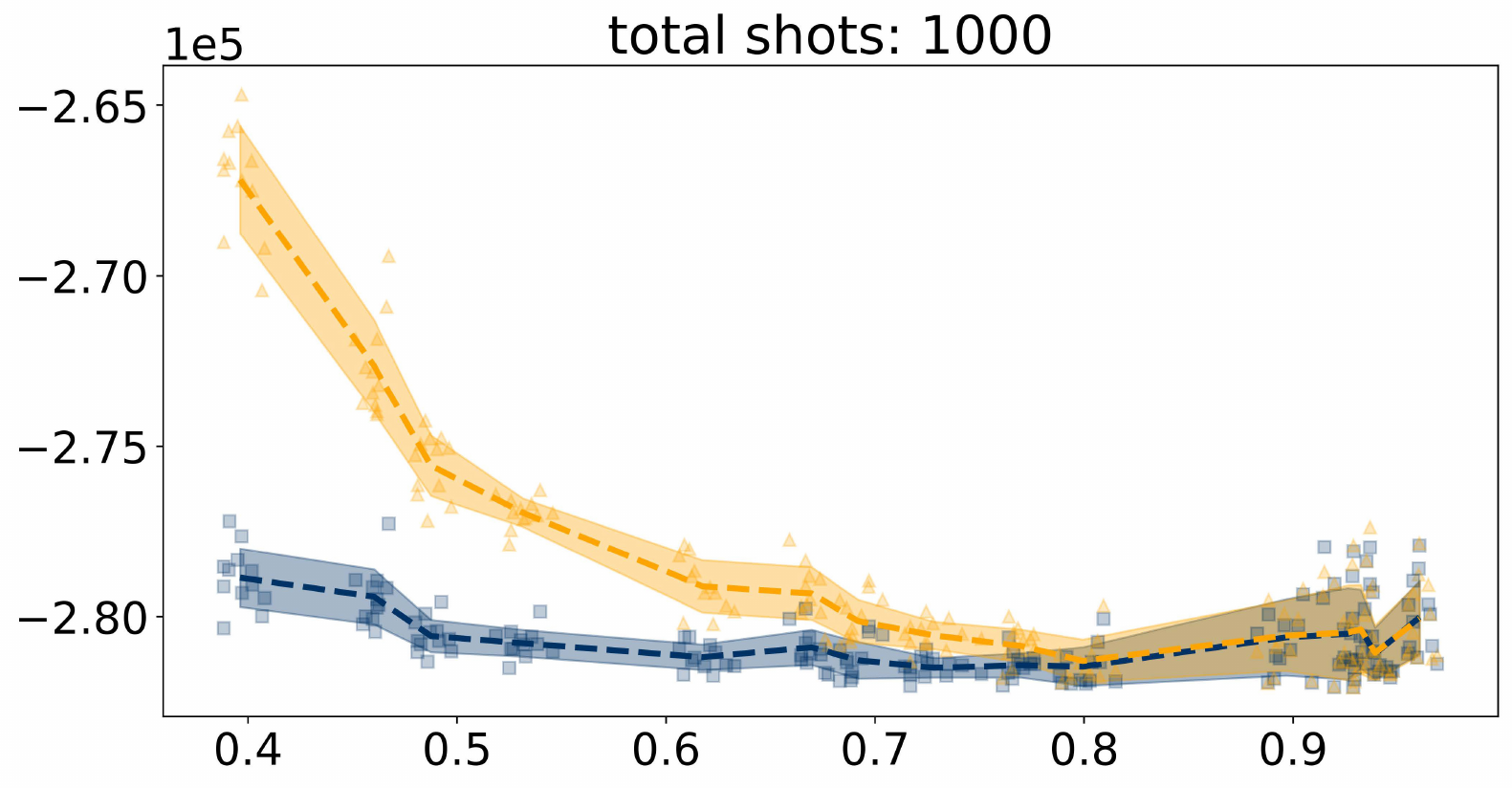}
         \label{subfig: spread 1000}
     \end{subfigure}
     \hfill
     \begin{subfigure}[t]{0.235\textwidth}
         \centering        \includegraphics[width=\linewidth]{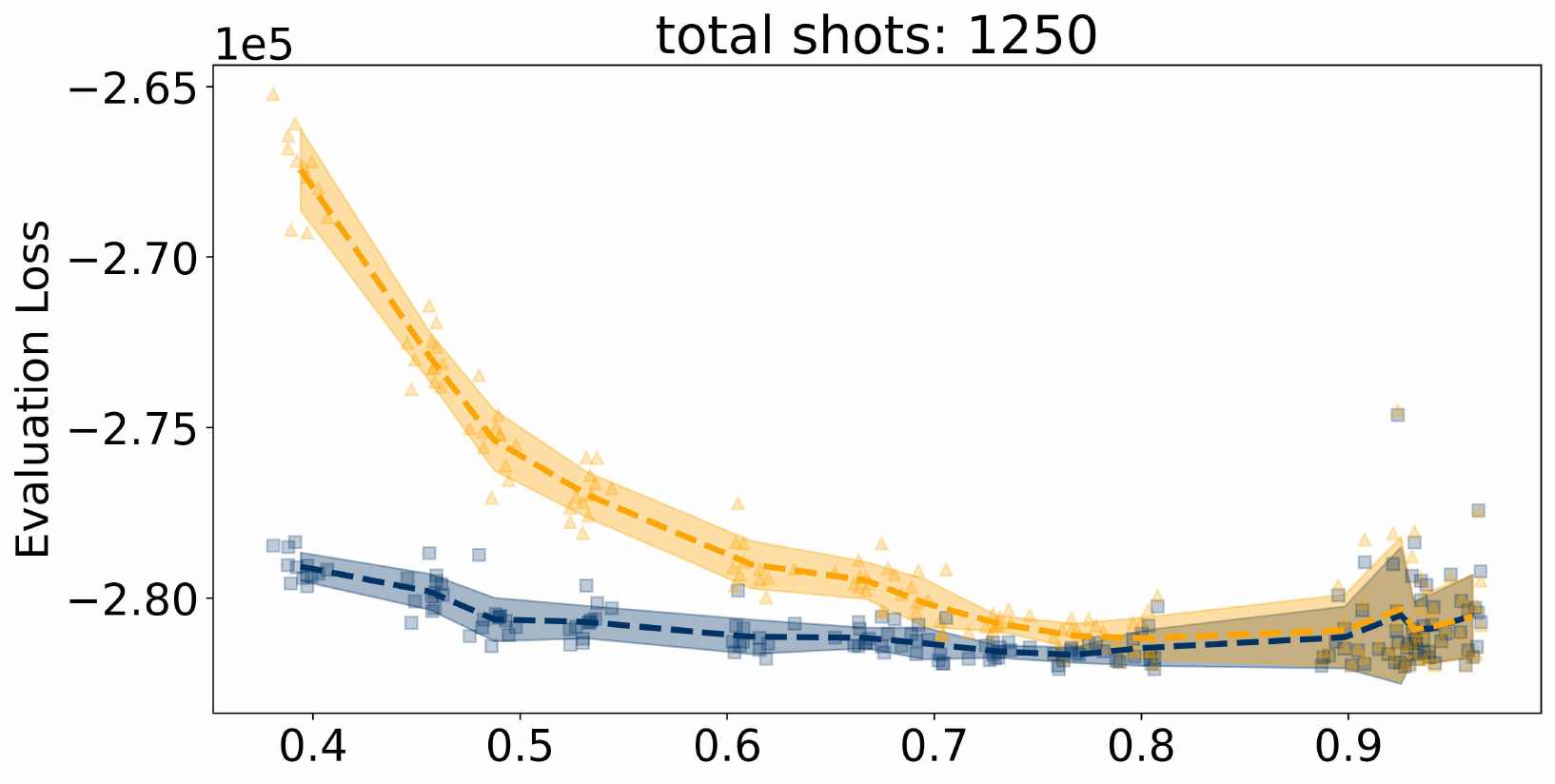}
         \label{subfig: spread 1250}
     \end{subfigure}
     \hfill
     \begin{subfigure}[t]{0.225\textwidth}
         \centering        \includegraphics[width=\linewidth]{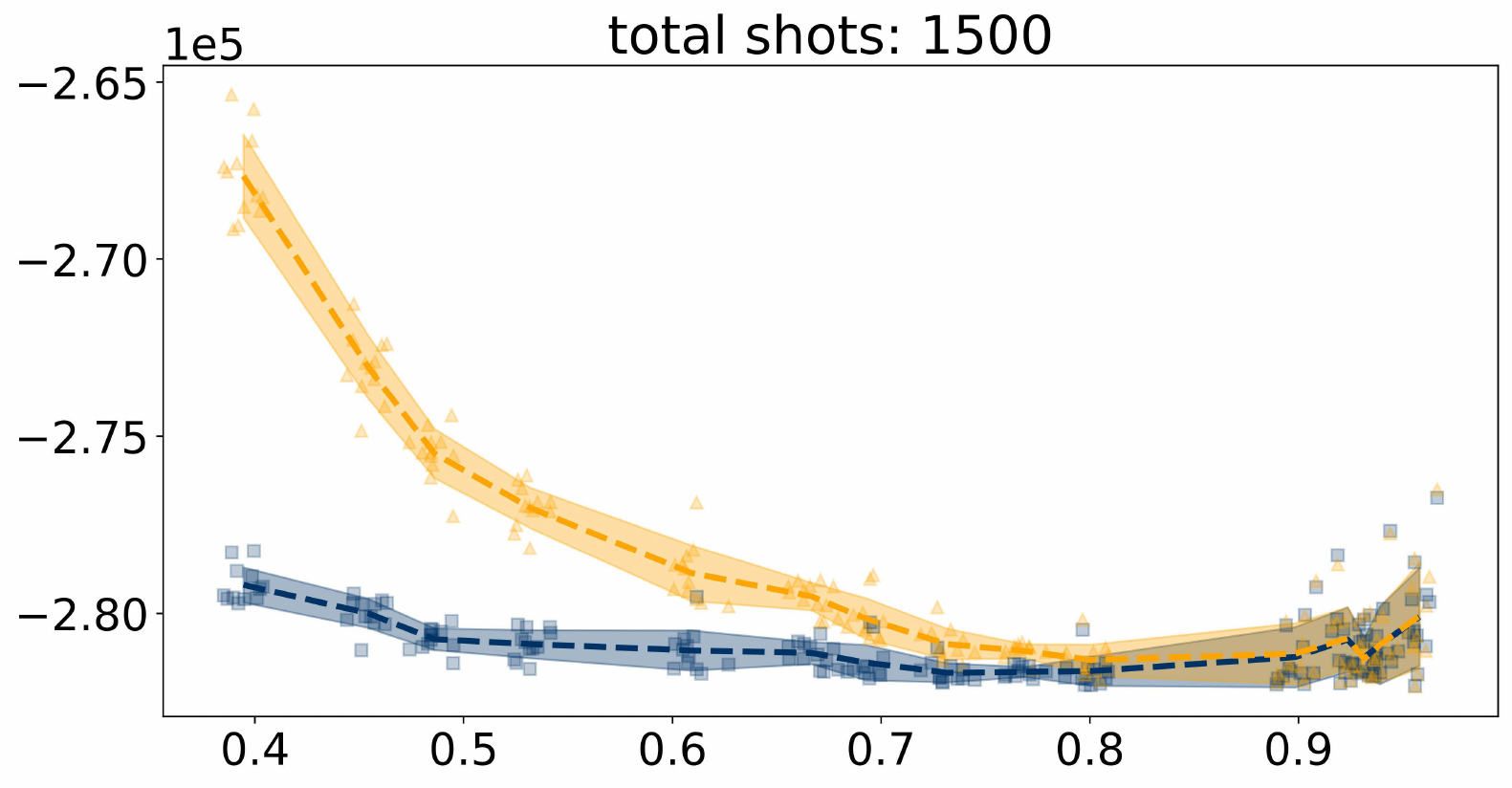}
         \label{subfig: spread 1500}
     \end{subfigure}
     \hfill
     \begin{subfigure}[t]{0.235\textwidth}
         \centering        \includegraphics[width=\linewidth]{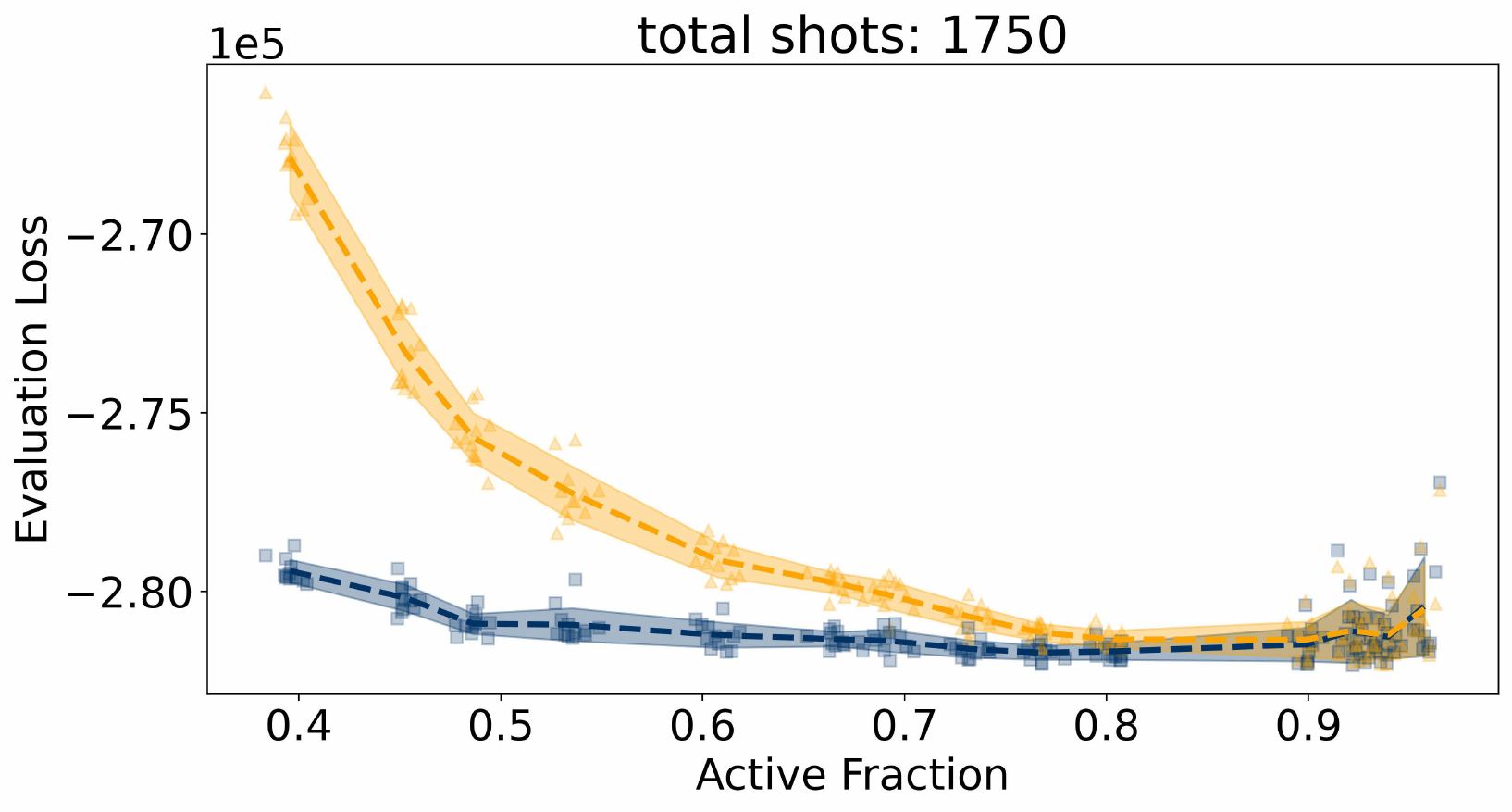}
         \label{subfig: spread 1750}
     \end{subfigure}
     \hfill
     \begin{subfigure}[t]{0.225\textwidth}
         \centering        \includegraphics[width=\linewidth]{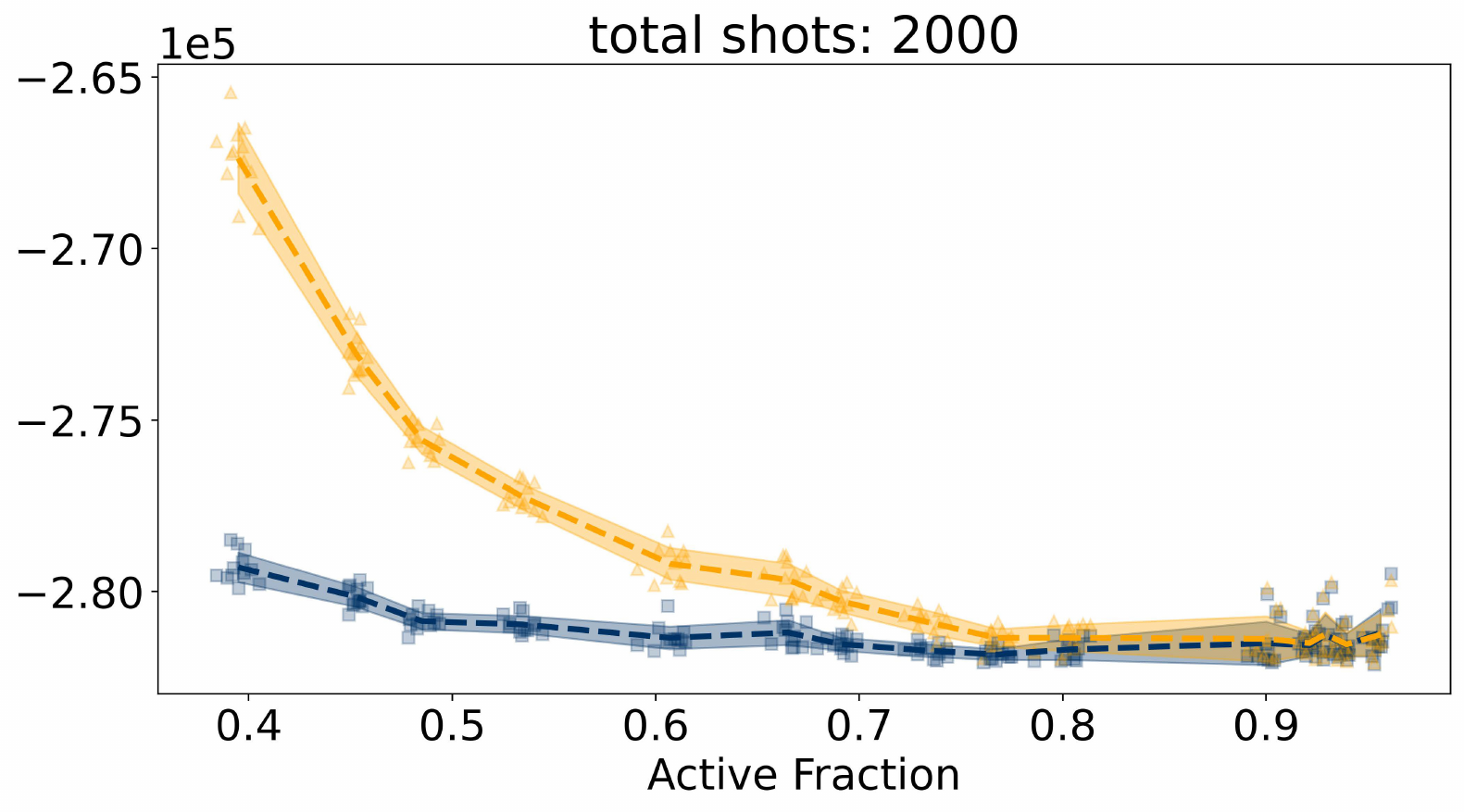}
         \label{subfig: spread 2000}
     \end{subfigure}
     
    \caption{The evaluation loss curves for 12 independent measurements of differing integration intervals and their means and variances were generated. Large variances indicate random error in the fit due to shot noise, while minor variances indicate minimal random error. The separation of the mean curves for each fitting routine demonstrates how deadtime bias is still the dominant source of error at high flux.}
    \label{fig: eval loss spread}
\end{figure}

The fitting routine was tested across different accumulation intervals (in time) to investigate the impact of shot noise. This was done by applying the deadtime- and Poisson-fitting routines to the measurements discussed in Sec. \ref{subsubsec: exp discussion}, but only using a fraction of the laser shots for this study. Each data superset was divided into 12 independent subsets by reducing the total shot counts per accumulation, i.e., creating 12 disjoint subsets (each composed of $10^3$ laser shots) from a superset (originally accumulated over $10^5$ laser shots). Decreasing the number of shots per subset would produce increasingly noisy histograms. Fits were generated for each subset with their evaluation losses calculated and plotted in Fig. \ref{fig: eval loss spread}. The distributions and mean values of the evaluation-loss curves were also calculated and shown for each superset. These statistics helped differentiate where fit quality was primarily influenced by random noise or deadtime bias. 

For context, the laser repetition rate was 14.3 kHz, so the lowest integration length of 250 laser shots corresponded to a short accumulation interval of approximately 17 ms while simultaneously using ultra-fine bin widths of 25 ps for acquisition. On the other end, the most prolonged interval was 136 ms for 2000 shots. Note that the range resolution selected in this study (25 ps or 4 mm) is much smaller than those typically used when processing atmospheric lidar measurements. These resolutions were chosen because they were small enough to generate the noisy fits needed for this investigation while ensuring that the error in the fits would decrease as bin widths are scaled up (to more typical values encountered in atmospheric lidar).

For low fluxes (AF $>0.7$) across all integration intervals, the mean values were similar between Poisson and deadtime models, while the standard deviations for both models were the largest, demonstrating that the variance in fit quality was primarily a consequence of random error. Conversely, at high flux (AF $<0.7$), the distribution spread decreased, indicating that shot noise was not affecting the fit quality as much. As the number of integrated shots increased, the spread in the distributions diminished, demonstrating that the random error in the fits was due to shot noise. Meanwhile, the mean Poisson curve grew in magnitude at high flux (or low AF) while the deadtime curve stayed relatively constant over the same interval. 

In these plots, the separation points for the mean Poisson and deadtime curves indicates where the deadtime model outperforms the Poisson model. For the largest integration interval (2000 shots), the separation point is distinct and occurs near AF$=0.8$. For the shortest integration interval (250 shots), the separation point is less distinct (due to the variance of the solutions) yet still occurs close to AF$=0.8$. These results show that the primary limitation of the deadtime model is the variance in the solutions due to shot noise. This analysis showed that the approach was still very effective in the presence of shot noise and only improved as the shot noise decreased (or accumulation length increased). It suggests that the deadtime noise model can effectively account for biases imposed by deadtime even when photon counts are sparse.  This represents a key enabling feature for applying MLE to sparse photon counting datasets.  Thus, these results demonstrate the robustness and agility of the estimator and suggest its potential for dynamic, transient, or ephemeral features, which are challenging measurements in atmospheric lidar. 

\subsubsection{Extended Signals} \label{subsubsec: extended signals}
The previous experiments demonstrated the deadtime model's performance at high range and temporal resolution on fine scale targets, representing key improvements over contemporary methods. The model not only applies to narrow pulses but is generalizable to all signal shapes. This is important because most atmospheric signals result from distributed scattering in a continuum (e.g., molecules and particles). This means that deadtime can occur randomly within the scattering volume and will often be shorter than the length of the target itself. As a result, it was important to validate the deadtime model on extended signals (which will be defined as having a temporal duration $\Delta t_{ext}$ greater than the cumulative deadtime, $\tau <\Delta t_{ext}$). In order to demonstrate the applicability of the deadtime noise model to atmospheric lidar, it was evaluated on extended signals that significantly exceed the deadtime interval. The tests were conducted using extended signals generated from laboratory experiments instead of real atmospheric data. The laboratory experiments offered a rigorous, controlled validation method that guarantees the correction adresses errors in deadtime, not compensating for other confounding factors, as would be the case if the method was applied directly to atmospheric lidar data. 

\begin{figure}[h!]
     \centering
     \begin{subfigure}[h]{0.475\textwidth}
         \centering        
         \includegraphics[width=1\linewidth]{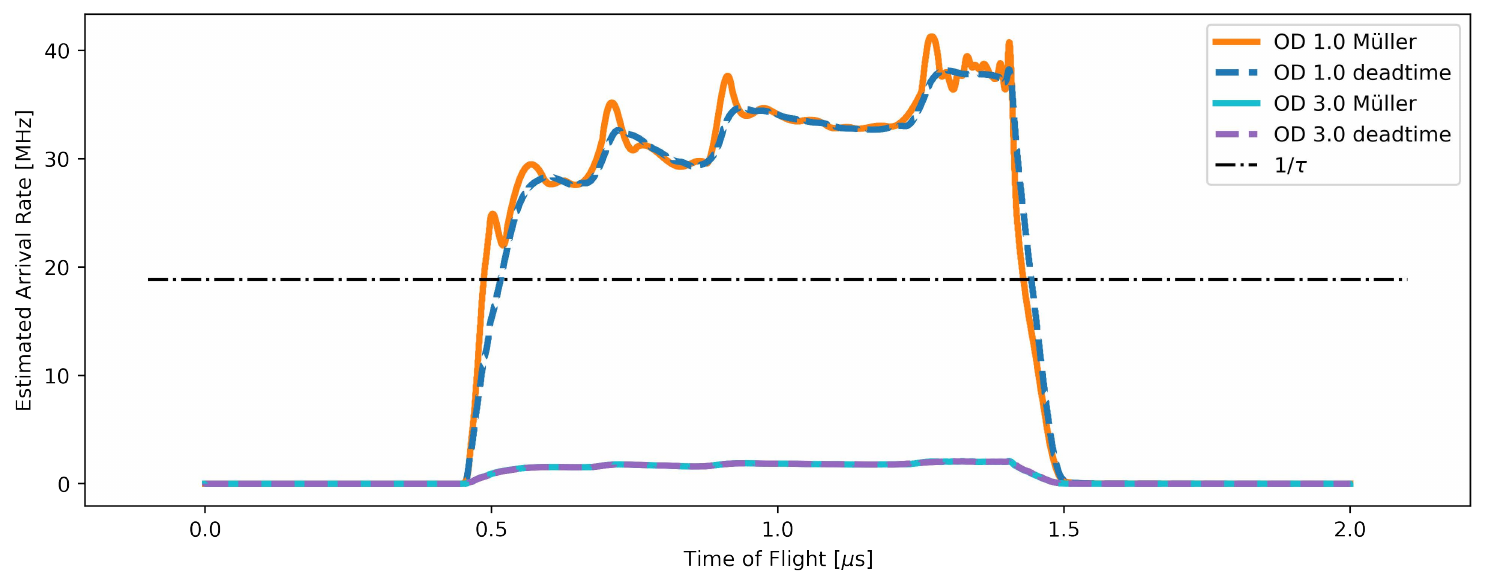}
     \end{subfigure}
     \hfill
     \begin{subfigure}[b]{0.475\textwidth}
         \centering      \includegraphics[width=1\linewidth]{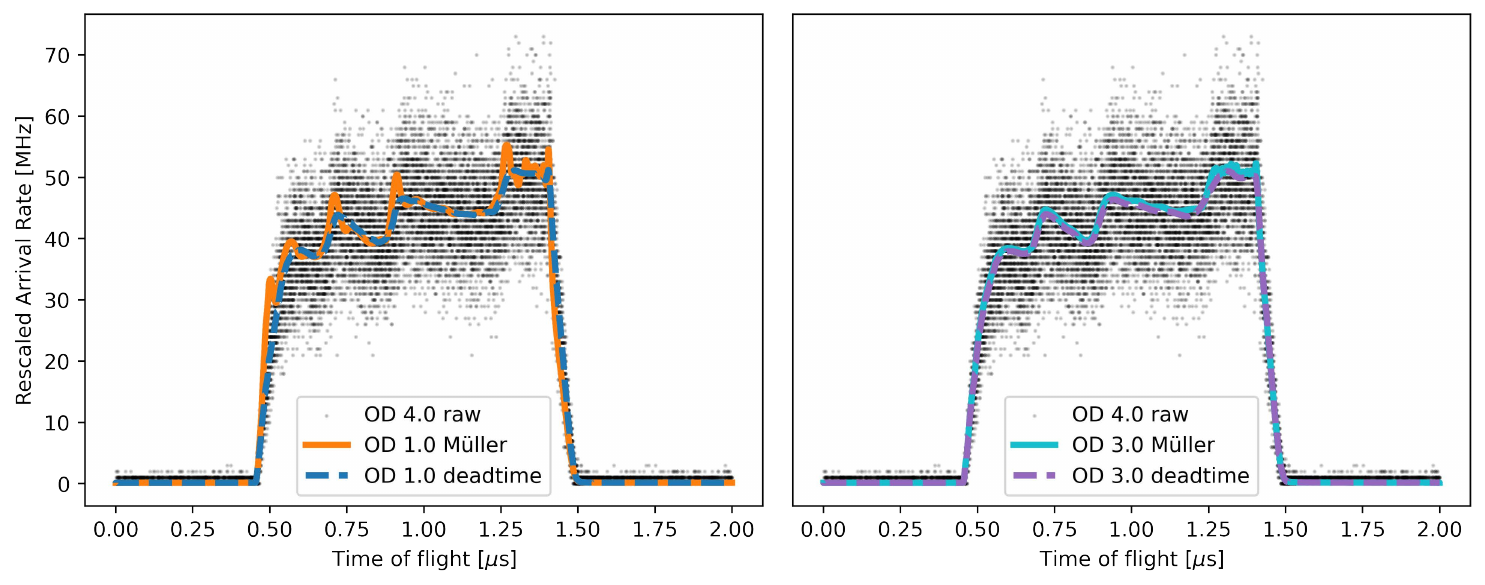}
     \end{subfigure}
     \caption{(Top) Comparisons of the retrieved curves using the deadtime noise model and the M\"uller correction for the OD 1.0 and 3.0 cases. The photon fluxes for the OD 1.0 case exceed the 1/$\tau$ flux threshold, while the OD 3.0 case fluxes do not. The plotted curves are limited to measurements using OD 1.0 and higher because the M\"uller correction fails when applied to data at higher flux (OD 0.0 and 0.3) due to the correction predicting negative photon counts. These higher flux cases were still processed using the deadtime model (see Fig. \ref{fig: extended target eval loss}). (Bottom) The same estimates as the top panel, but the recovered flux was rescaled to optimally match the linear OD 4.0 data, which is shown in the gray dots. The panel is divided into two by OD to improve visual comparison of the models. When there was little deadtime bias (right), both corrections recover the signal without bias. When there was large deadtime bias (left), the M\"uller-corrected fit imparts substantial bias.}
     \label{fig: extended target curves}
 \end{figure}

 Like the previous effort, the experiment used a TCSPC system to interrogate scattering from a laser pulse with varying attenuation levels. In this case, the laser pulse was generated using the NSF National Center for Atmospheric Research MicroPulse DIAL laser source, a distributed Bragg reflector diode laser coupled with a tapered semiconductor optical amplifier \cite{MPD}. The pulses were approximately 1 $\mu$s long and contained structure due to interference from varying amplifier modes, an adequate proxy for atmospheric structure. The output pulse was measured using the SPCM detector with a deadtime interval of 53 ns (note that the pulse length was longer than the detector deadtime by a factor of 20). An ND filter was inserted in the laser path to vary signal flux with OD values of 0.0, 0.3, 1.0, 3.0, and 4.0. The data was collected over $3.2\times10^6$ laser shots for each OD value. Note that the filter's transmissivity $T_{\mathrm{ND}}$ drops exponentially as a function of the OD value according to $T_{\mathrm{ND}}=10^{-\mathrm{OD}}$. The data from the OD 4.0 case served as the linear evaluation dataset since it was largely unaffected by deadtime. The method for evaluating the recovered signal is thus nearly identical to the previous experiments, where a low-flux (limited effect of deadtime) evaluation dataset was used to assess recovered signal quality. 

In each attenuation case, the photon flux of the attenuated laser pulse was estimated using maximum-likelihood estimation based on the noise model under evaluation (deadtime or Poisson with M\"{u}ller-corrected counts). In this case, the laser pulse was not as well represented by smoothly varying signals described by exponential polynomials. Instead, splines with varying knot spacing as the basis functions was used for the maximum-likelihood estimate. The knot spacing of the splines was determined using dyadic tree trimming with holdout cross validation (the knot configuration was tuned in the same way as the polynomial order in the previous analysis). Note that in the extended target case, the M\"{u}ller correction's violation of Approximation \#2 was much less significant and thus allows us to use it as a reasonable baseline for evaluating the deadtime noise model.

\begin{figure}[t]
    \centering    \includegraphics[width=0.375\textwidth]{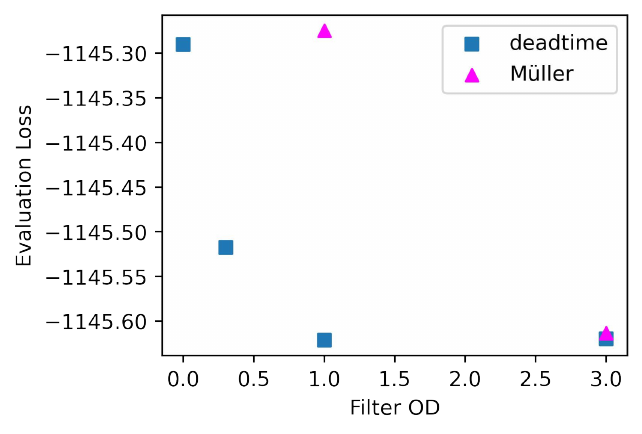}
    \caption{The evaluation loss scores for the M\"uller correction and deadtime models are not included for the M\"uller correction at OD 0.3 and OD 0.0 because the correction fails at these higher fluxes. The deadtime model's evaluation loss was lower (or better) than the M\"uller correction across attenuation levels.}
    \label{fig: extended target eval loss}
\end{figure}

The top panel of Fig. \ref{fig: extended target curves} shows the retrieved photon flux for the OD 1.0 and 3.0 cases. Note that the flux in the OD 1.0 case exceeded the 1/$\tau$ rate by nearly a factor of two (meaning significant deadtime bias), while the flux in the OD 3.0 case was approximately a factor of four below this value (meaning little deadtime bias). In the OD 3.0 case, both the retrievals agree relatively well with the OD 4.0 data and with each other.  The retrieval using the deadtime noise model only slightly outperforms the M\"{u}ller-corrected fit on the evaluation negative log likelihood (see Fig. \ref{fig: extended target eval loss}).  In the OD 1.0 case, the M\"{u}ller-corrected fit produced noticeable artifacts in the retrieval, the most obvious cases corresponding to regions where the photon flux changed rapidy. At the same time, the retrieval using the deadtime noise model for the OD 1.0 case accurately recovered the same pulse features seen at lower fluxes in the OD 3.0 case (even outperforming the M\"{u}ller correction at OD 1.0). These results suggest that when the M\"{u}ller correction is applied to high-flux atmospheric signals with similar structure (e.g., clouds and dense aerosol layers), the method imparts similar systematic biases. This not only reduces the accuracy of the retrieval but would likely result in misinterpretation of the target structure, i.e., variability becomes exaggerated. The deadtime model does not introduce this bias, meaning that atmospheric data corrected using the deadtime model would exhibit less bias and introduce fewer artificial features. 

Fig. \ref{fig: extended target eval loss} displays the evaluation loss scores that compare the deadtime and M\"{u}ller-corrected fits to the OD 4.0 (linear) evaluation dataset.  This analysis is identical to that performed in the experiments presented earlier in this work. The results show that the deadtime model always produced a better (or lower) evaluation score than the corresponding M\"{u}ller-corrected estimates. Despite the very high flux associated with the OD 1.0 case (estimated peak flux of 40 MHz), the deadtime noise model still produced a lower evaluation loss than the M\"{u}ller correction for the OD 3.0 case (estimated peak flux of 2 MHz), extending dynamic range by over one order of magnitude. This extension was nearly matched even at OD 0.0 (no attenuation with estimated peak flux of 250 MHz), where the deadtime correction still outperformed the M\"{u}ller correction at OD 1.0. These results demonstrate that the deadtime noise model generalizes to high-flux extended signals that are characteristic of volumetric targets in the atmosphere. This means that the deadtime noise model far exceeds any existing approach to correct for deadtime in atmospheric lidar. 

It is important to emphasize that the targets interrogated here are static targets, a key factor in the flux estimation performance and establishing the effectiveness of the deadtime noise model.  As high-flux atmospheric targets also tend to be temporally dynamic, the work presented here represents just one step in addressing these complex targets and enables future work investigating other factors influcencing lidar data quality.

\section{Conclusion}

This study presented a photon-counting noise model that encapsulates deadtime effects for processing photon-counting lidar data, thus enabling accurate recovery of photon flux from SPAD detectors close to their saturation point. It was first demonstrated that the traditional deadtime correction approach (or the M\"{u}ller Correction) is limited to scenes that are (1) constant in flux and (2) acquired at coarse resolution with respect to the detector deadtime. This limits its utility for deadtime correction in atmospheric lidar where atmospheric features do not always follow these restrictions. Then, the new deadtime noise model was derived and evaluated using measurements that violated these restrictions (i.e., variable flux, high resolution). By employing a maximum-likelihood estimator with the novel deadtime noise model, the dynamic range of flux measurements was extended beyond the traditional approach. This was tested in simulation and experiment by estimating narrow targets across fluxes that spanned multiple orders of magnitude. The results confirmed that the deadtime noise model enables high-resolution measurements and extends dynamic range by over an order of magnitude beyond contemporary methods. The estimator was also successful in the presence of shot noise, demonstrating it as a candidate noise model for processing sparsely distributed photon-counting data, such as in Ref. \cite{Hayman-2D-estimation}. These results support more accurate observations of high-flux targets that exhibit spatial and temporal variability (e.g., cloud edges, smoke plumes), creating a new pathway to understand their underlying processes. The results also suggest its benefit for additional applications containing sparse data, such as transient or tenuous features, optically thick targets, and dynamic observational platforms. 

Finally, the results showed that the model generalizes to extended targets typical in atmospheric lidar, extending the dynamic range well beyond the 1/deadtime limit. This work provides a foundation for further study into temporally dynamic atmospheric targets by firmly establishing and validating a mechanism correcting errors imparted by deadtime.  By leveraging this approach, it should be possible to understand errors imposed by high flux and temporal dynamics and develop sensor and processing strategies to mitigate them.   

% \begin{backmatter}

\section*{Acknowledgements}

The material is based upon work supported by the National Aeronautics and Space Administration (NASA) Federal Award No. 80NSSC22K1212; the National Science Foundation (NSF) National Center for Atmospheric Research, which is a major facility sponsored by the NSF under Cooperative Agreement No. 1852977; NSF Grant No. AGS-1930907; and NASA Earth Science Technology Office Decadal Survey Planetary Boundary Layer Incubation Program. 

This work was supported by a NASA Space Technology Graduate Research Opportunities (NSTGRO) fellowship.  The lead author would like to acknowledge Amin Nehrir and Xiaoli Sun for their guidance and support through the NSTGRO research fellowship. The authors would also like to thank Robert Stillwell, Kevin Sacca, and Alexandra Wise for their support during this work. 

\section*{Disclosures}
The authors disclose no conflict of interest.

% References
\printbibliography

\section{Appendix} \label{sec: appendix}

\renewcommand{\thesubsection}{\Alph{subsection}}

\subsection{Violation of M\"uller Approximation \#2} \label{subsec: Muller appendix}
\begin{figure}[h!]
\centering
\begin{subfigure}[t]{0.42\textwidth}
\centering
\includegraphics[width=\textwidth]{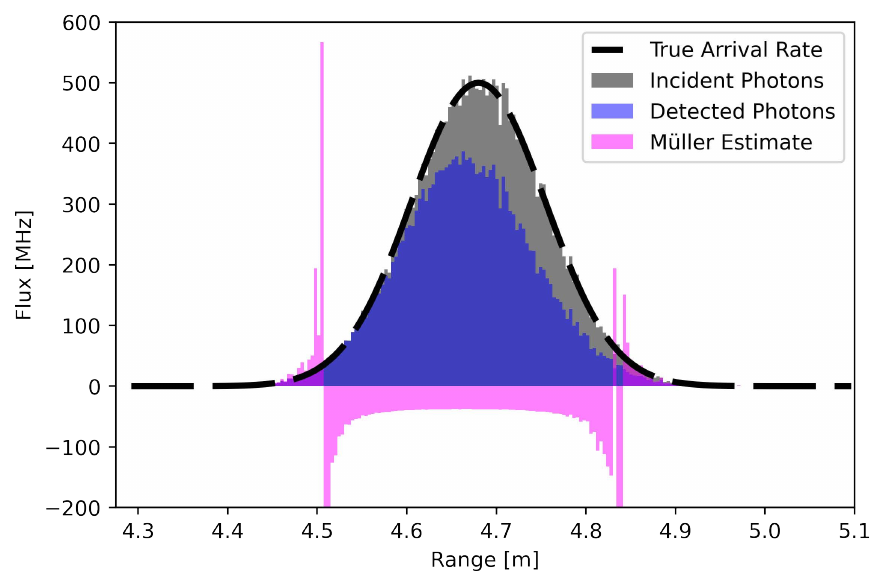}
         \caption{}\label{subfig: Muller Nonphysical}
\end{subfigure}

\begin{subfigure}[t]{0.42\textwidth}
\centering
\includegraphics[width=\textwidth]{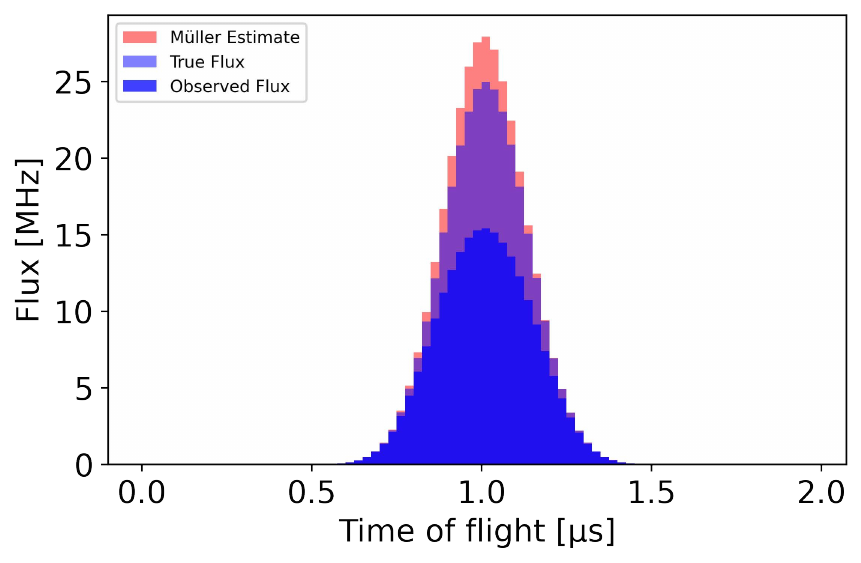}
         \caption{}\label{subfig: Muller Physical}
\end{subfigure}
    \caption{Examples where the M\"uller Correction was applied at resolutions that violated M\"uller Approximation \#2. (Top) A severe example where the solution is non-physical. (Bottom) Inaccurate M\"uller estimate when applied to the profile in Fig. \ref{subfig: Muller histograms}.} 
    \label{fig: Muller correction test at high resolution}
\end{figure}

Fig. \ref{fig: Muller correction test at high resolution} shows examples where the M\"uller Correction was applied to profiles at high resolutions that violated M\"uller Approximation \#2. In Fig. \ref{subfig: Muller Nonphysical}, the instantaneous fluxes were high enough that the M\"uller estimate produced a nonphysical result, which is a consequence of applying the M\"uller Correction at high resolution. In Fig. \ref{subfig: Muller Physical}, the solution was also inaccurate but can be interpreted as physical, which highlights how the application of the M\"uller Correction at high resolutions like this can result in inaccuracies that pass undetected without careful inspection.

\subsection{Evaluation Loss Scale Factor}  \label{subsec: eval loss scale factor}

The forward model output from the fitting routine $\tilde{\lambda}(t)$ must be scaled optimally by the factor $T_r^*$ for proper comparison with the evaluation dataset. This relationship is given by
\begin{equation}
    \tilde{\lambda}^*(t)=T_r^*\tilde{\lambda}(t)
\end{equation}
This is done by applying the evaluation dataset to the Poisson loss function (which is valid for the evaluation dataset which was acquired over very low flux):
\begin{equation}
    \mathcal{L}_{eval}=\sum_{m=1}^M\left(\mathcal{N}\tilde{\lambda}_{m}^*\Delta t-Y_{m}^{(e)}\ln(\tilde{\lambda}_{m}^*)\right)
\end{equation}
where $\tilde{\lambda}_{m}^*\triangleq\tilde{\lambda}^*(t_m)$. The optimal scaling factor is then found by minimizing the loss function:
\begin{equation}
    \frac{\partial \mathcal{L}_{eval}}{\partial T_r^*} = 0 = \sum_{m=1}^M\left(\mathcal{N}\tilde{\lambda}_m\Delta t - \frac{1}{T_r^*}Y_{m}^{(e)}\right)
\end{equation}
\begin{equation}
    T_r^* = \frac{1}{\mathcal{N}\Delta t}\frac{\sum_{m=1}^MY_{m}^{(e)}}{\sum_{m=1}^M\tilde{\lambda}_m}
\end{equation}

\end{document}